\documentclass{article}       
\usepackage{graphicx}
\usepackage{color,verbatim}
\usepackage{lscape}
\usepackage{rotating}
\usepackage{wrapfig}
\usepackage{multirow}
\usepackage{placeins}
\usepackage{amsmath,bm}
\usepackage{amsfonts}

\usepackage{lmodern}
\usepackage{microtype,color}
\usepackage[numbers,square,sort&compress]{natbib}

\usepackage{bbm}
\usepackage{mathtools}
\usepackage{amsmath}
\usepackage{float}
\usepackage{subcaption}

\usepackage{xcolor}
\usepackage[algo2e,ruled]{algorithm2e}

\begin{document}

\title{Bayesian inference for optimal dynamic treatment regimes in practice}

\author{Daniel Rodriguez Duque, Erica E.M.~Moodie, and David A.~Stephens}

\maketitle

\begin{abstract}
In this work, we examine recently developed methods for Bayesian inference of optimal dynamic treatment regimes (DTRs). DTRs are a set of treatment decision rules aimed at tailoring patient care to patient-specific characteristics, thereby falling within the realm of precision medicine. In this field, researchers seek to tailor therapy with the intention of improving health outcomes; therefore, they are most interested in identifying \textit{optimal} DTRs. Recent work has developed Bayesian methods for identifying optimal DTRs in a family indexed by $ \psi $ via Bayesian dynamic marginal structural models (MSMs) \cite{RodriguezDuque2022}; we review the proposed estimation procedure and illustrate its use via the new \texttt{BayesDTR} \texttt{R} package. Although methods in \cite{RodriguezDuque2022} can estimate optimal DTRs well, they may lead to biased estimators when the model for the expected outcome if everyone in a population were to follow a given treatment strategy, known as a value function, is misspecified or when a grid search for the optimum is employed.  We describe recent work that uses a Gaussian process ($\mathcal{GP}$) prior on the value function as a means to robustly identify optimal DTRs \cite{RodriguezDuque2022b}. We demonstrate how a $ \mathcal{GP} $ approach may be implemented with the \texttt{BayesDTR} package and contrast it with other value-search approaches to identifying optimal DTRs. We use data from an HIV therapeutic trial in order to illustrate a standard analysis with these methods, using both the original observed trial data and an additional simulated component to showcase a longitudinal (two-stage DTR) analysis. \\ \ \\

\end{abstract}

\section{Introduction}

Precision medicine builds on the concept of evidence-based medicine to determine not just the average efficacy of therapeutic or surgical interventions, but which intervention is right for whom. With this aim in mind, statisticians have sought to develop methods that allow for the discovery of tailored interventions. This has been done via statistical methods for dynamic treatment regimes (DTRs). DTRs are a set of decision rules that take patient information as inputs and that output a decision \cite{Murphy2001}. Most importantly, researchers in this realm have proposed methods that can determine the causal effect of being assigned to a specific DTR and to identify optimal DTRs, that is, DTRs with the highest expected outcome, or value. Frequentist inference has been given much attention to this field while Bayesian methods have received significantly less heed. In this work, we examine methods that allow for  Bayesian causal inference of optimal DTRs, in particular methods that can robustly identify the optimal strategy.

There are many frequentist methods for identifying optimal DTRs. These include, though are not limited to, g-computation \citep{Robins1986}, g-estimation of structural nested models \citep{Robins1993},  Q-learning \citep{zhao2009}, dynamic marginal stuctural models (MSMs) \citep{Orellana2010}, and outcome weighted learning \citep{zhao2012}. Bayesian methods have also been proposed, including by Saarela et al. \cite{Saarela2015a} who use a predictive Bayesian approach that requires the specification of parametric distributions for outcomes and intermediate covariates, Murray et al. \cite{murray2018} who propose a Bayesian adaptation to Q-learning, Arjas et al. \cite{Arjas2010} who use Bayesian nonparametric regression and backward induction, and Xu et al. \cite{Xu2016} who use Bayesian nonparametrics in a survival context, where patients can transition between disease states. Hua et al. \cite{hua2022personalized} address the question of identifying optimal treatments, in addition to optimal treatment times, by proposing a Bayesian joint model for the sequence of medical interventions and for the clinical measurements, including intermediary covariates and the final outcome. Recently, a Bayesian method for inferring optimal DTRs via dynamic MSMs was developed \cite{RodriguezDuque2022}. In addition to allowing for population-level inference, this approach also allows for individualized inference by enabling a decision-maker to determine whether a patient with a specific set of characteristics is receiving optimal therapy. An overview of Bayesian DTRs can be found in \cite{oganisian2021practical}.

Although Bayesian inference via dynamic MSMs enables identifying optimal DTRs, limitations remain; for example, inference hinges on the correct specification of a marginal model. A Gaussian Process ($ \mathcal{GP} $) prior has recently been proposed to model the value function and consequently identify optimal DTRs via a sequential sampling scheme \cite{RodriguezDuque2022b}. In principle, $ \mathcal{GP} $-based methods can utilize any estimator for the expected outcome under adherence to a DTR, known as the value for the regime, and avoid drawbacks associated with some value-search approaches. For example using a dynamic MSM to directly model the value surface may not perform well if the model is misspecified. Alternatively, if a grid search is used to obviate the issues of directly modeling the value surface, an inefficient procedure results which may incorrectly identify the optimal DTR when the value function is multi-modal and which may be computationally intractable when Bayesian estimators are utilized. A Bayesian approach that uses $ \mathcal{GP} $s to represent uncertainty in the value function  has the potential to more efficiently utilize information by selecting experimental points that are expected to be optimizers and by providing a very flexible model for the value function.

In this paper, we aim to review how a Bayesian approach may capitalize on semiparametric inference as presented in \cite{Saarela2015} and \cite{RodriguezDuque2022} in order to identify optimal DTRs. This is important as the ideas required for this inferential approach are nuanced and therefore challenging for practitioners to implement. We introduce a new package \texttt{BayesDTR} to illustrate how to utilize these methods in practice. With these foundations in place, we further study how $ \mathcal{GP} $ optimization can be used to identify optimal DTRs, and we examine how the \texttt{BayesDTR} package provides functionalities to perform an analysis reliant on these methods. There are several packages available in the Comprehensive \texttt{R} Archive Network (CRAN) that performs estimation or inference about DTRs. These include \texttt{DTRreg} which implements dynamic weighted least squares, g-estimation, and Q-learning \citep{Wallace2020}, \texttt{DTRlearn2} which performs outcome weighted learning \citep{Chen2020}, \texttt{DynTxRegime} which permits several methods including inverse probability weighting (IPW) and augmented IPW \citep{Holloway2020}, and \texttt{SMARTbayesR} which allows for Bayesian inference of optimal DTRs with data arising from SMART designs with binary outcomes \citep{Artman2021}. Currently, there are no packages that allow for Bayesian semiparametric inference of optimal DTRs, nor any that directly use $ \mathcal{GP} $ optimization with estimators for the value of a DTR. Roustant et al. \cite{Roustant2012} have developed a package, \texttt{DiceOptim}, for optimization using $ \mathcal{GP} $ methods, though this package is not tailored to estimators for the value function nor does it address how to quantify uncertainty around the optimum.

This manuscript is organized as follows: section 2 introduces recently developed Bayesian methods for identifying optimal DTRs, section 3 describes the functions in the \texttt{BayesDTR} package that allow for the use of these methods. For illustrative purposes, we adapt data from the ACTG175 (AIDS Clinical Trials Group Study 175) trial, available in the \texttt{LongCART} \texttt{R} package \cite{Kundu2021}, to perform a plasmode simulation depicting a standard analysis with these methods and package. Section 4 demonstrates how to use this package to perform a standard analysis with these methods. We summarize and conclude in section 5.

\section{Methods}

\subsection{Bayesian Dynamic MSMs for Optimal Dynamic Regimes}
In this section, we examine how to perform inference for optimal DTRs via dynamic MSMs, using the methods developed by \cite{RodriguezDuque2022}. To do this, the inferential setting must first be formalized and notation defined. Consider a multi-stage decision problem with $K$ decision points and final continuous-valued outcome $ y $. At every decision point $ k $, a set of covariates $ x_k $ is observed. It is assumed that these consist of all time-fixed and time-varying confounders, if there are any. Covariate history up to time $ k $ is denoted by $\bar{x}_k=\{x_1,...,x_k\}$, and observed treatment history up to stage $ k  $ is given by $\bar{z}_k=(z_1,...,z_k)$, $z_j \in \{0,1\}$. Subscripts are omitted when referencing history through stage $  K $. All patient information is grouped into $b=(\bar{x},\bar{z},y)$. As interest is centered around examining the effect of adherence to specific DTRs, the DTR-enforced treatment history can be considered by $g(\bar{x})=(g_1(x_1),...,g_K(\bar{x}_K)), \; g_i(\bar{x}_i) \in \{0,1\}$. This is the sequence of treatments that would be observed if a patient followed a treatment strategy $ g $ throughout the entire follow-up period; it contrasts the treatment history $ \bar{z} $ that is observed in patients in an analytic dataset. The observed treatment history $ \bar{z} $ and the DTR-enforced treatment history $g(\bar{x})$ only coincide in patients who have treatments consistent with those suggested by a DTR $ g $. The DTR-enforced treatment history up to stage $ k $ is given by $ \bar{g}_k(\bar{x}_k)=(g_1(x_1),...,g_k(\bar{x}_k)) $. {Attention is restricted to a family $ \mathcal{I} $ of DTRs indexed by  $\psi \in \mathcal{I}$ to give $ \mathcal{G}=\{g^\psi(\bar{x}); \psi \in \mathcal{I}\}$. This family $ \mathcal{I} $ can have different dimensions depending on the problem of interest, for example treatment rules of the from "treat when $ x_t>\psi, \psi \in (0,1)$" lead to a family containing a continuum of $ \psi $s. Note that we only consider deterministic DTRs that assign treatment deterministically using patient information. An example of a DTR indexed by a parameter $ \psi $ is one of the form "treat at stage $ k $ when $ x_k>\psi_k $". Interest lies in two treatment and covariate distributions: the \emph{observational} world distribution $P_\mathcal{O}$ which denotes the law giving rise to the data in the study population, and the \emph{experimental} world distribution $P_\mathcal{E}$, which is problem specific, and should be defined such that causality can be inferred. Under these two worlds, the marginal distribution of $ x_1 $ is identical, and the dependence of $ x_k $ on previous treatment and covariates is also unchanged for $ k=2,...,K $, however the component of the joint distribution governing treatment allocation differs. For example, Saarela et al. consider a world in which treatments are sequentially randomized so that stage-specific treatment effects can be estimated \cite{Saarela2015}. Rodriguez Duque et al. focus on an experimental world where patients are randomly assigned to a DTR in $ \mathcal{I} $ at study start \cite{RodriguezDuque2022}. Lastly, variables sampled from a posterior distributions are shown with $^*$.

Inference for Bayesian dynamic MSMs begins by considering a utility $ U(b,g^\psi,\beta) $, with $ \beta $ being a parameter that we can use to maximize the utility; focus is on the negative squared error loss utility, $U(b,g^\psi,\beta)=-(y-h(\beta,\psi))^2$, where $ h(\beta,\psi) $ models $ E_{\mathcal{E}}[Y|G=g^\psi] $, indexed by an unknown parameter $\beta$ and where the expectation is taken with respect to the true data-generating distribution in the experimental world. For this specific choice of utility, which aims to minimize the square distance between observed outcomes and their marginal means,  no other elements of $ b $ are required. This utility is of interest because it allows for an explicit model of the quantity of interest, the expected outcome under assignment to a regime $ g^\psi $, in a world where regime assignment is unconfounded. For a Bayesian decision-maker, interest lies in the value of $\beta$ that maximizes the posterior expected utility $E_{\mathcal{E}}[U(\bar{B}^*,G,\beta)|\bar{b}]$, where $ G $ is the random variable denoting regime assignment. The expectation taken with respect to the experimental measure in which patients are randomized to regimes in $\mathcal{G}$ at study start, with probability $p(G=g)$. The basis for this decision theoretic approach is well laid out in \cite{Walker2010a}. When a finite set of regimes is considered, with patients having equal probability of randomization, $ p(g) $ can be replaced  with $1/C_G$, where $C_G=|\mathcal{I}|$.
With a chosen utility, the next step in this approach lies in linking $\mathcal{E}$ and $\mathcal{O}$ with respect to a posterior predictive distribution. The required linkage is given by the following equation:
\begin{align}
	E_\mathcal{E}[U(B^{*},G,\beta)|\bar{b}]=  E_\mathcal{O}\left[ \frac{1}{C_G}\sum_{\{\psi\in \mathcal{I}\}} w^{\psi*}U(B^*, g^\psi,\beta)  \middle|\bar{b}\right],
	\label{eq:importancesamp}
\end{align}
with weight $w^{\psi*}$ given by
\begin{equation}
	w^{\psi*}=\frac{\mathbbm{1}_{g^\psi(\bar{X}^*)}(\bar{Z}^*)  }{\prod_{j=1}^{K}p_\mathcal{O}(Z_j^* |\bar{Z}_{j-1}^*,\bar{X}_{j}^*,\bar{b})}.
	\label{weight_equation}
\end{equation}
The denominator in the weight is the treatment probability in the observational world. The numerator is the probability of a sequence of treatments conditional on regime assignment; as only deterministic DTRs are considered, these probabilities are either 0 or 1, thereby yielding the indicator function. Randomization to regime $g^\psi$ is equiprobable for all regimes in the experimental world, and this is captured by the constant $C_G$. The $ ^* $ notation clarifies that the expectation in equation \eqref{eq:importancesamp} is taken with respect to a posterior predictive distribution. For equation \eqref{eq:importancesamp} to hold, a patient following regime $ g^\psi$ with recorded history $(\bar{x}_K, \bar{z}_{K})$ should have a positive probability of being observed in the observational world; effectively this is the positivity condition encountered in the causal inference literature \cite{Murphy2001}. Additionally, as is frequently found in the causal inference literature, the sequential no unmeasured confounders assumption is also required \cite{Murphy2001}. Note that the weight formula is \textit{not} stabilized and that having a stabilization term, meaning a term in the numerator containing the marginal treatment probabilities in $ \mathcal{O} $ \cite{Robins2000},  would change $ \mathcal{E} $ to one where the marginal treatment probabilities are as in $ \mathcal{O} $. It may be that the resulting probability law in $ \mathcal{E} $ is not well defined, given that treatments in $ \mathcal{E} $ are dictated by DTRs.

Having linked the experimental world with the observational world, focus becomes centered on how to infer about the parameters of interest $\beta$. Equation \eqref{eq:importancesamp} now allows for the use of observed world data to perform posterior inference in the experimental world. To perform Bayesian inference in this setting, a prior must be specified. Unlike parametric Bayesian inference, where a prior for $ \beta $ is specified directly, a prior is placed on the family of data generating distributions in the observational world $P_\mathcal{O}$,  denoted by $P_\mathcal{F}$. Effectively, this prior induces a prior on $\beta$ as $P_B(\beta \in \Omega)=P_\mathcal{F}(\{P_\mathcal{O}: \beta(P_\mathcal{O})\in \Omega\})$. The prior of choice is the nonparametric Dirichlet process $\mathcal{DP}(\alpha,G_x)$ prior  with scaling parameter $|\alpha| \to 0$. This prior has the benefit of converging asymptotically to the true data-generating distribution \cite{ghosal2017}. Under this specification, the Bayesian bootstrap yields the posterior predictive distribution \cite{Rubin1981}. A sample drawn from the posterior $\mathcal{DP}$ is given by $p_\mathcal{O}(b^*|\bar{b},\pi)=\sum_{i=1}^n \pi_i \mathbbm{1}_{b_i}(b^*)$, where $\pi=(\pi_1,...,\pi_n)$ is a sample from $\pi \sim Dir(1,...,1)$, a Dirichlet distributed random variable with all concentration parameters equal to one. Under the $ \mathcal{DP} $ prior that yields the Bayesian bootstrap, any distribution sampled from the posterior  $ \mathcal{DP} $ is uniquely determined by $\pi$. Stephens et al. \cite{Stephens2021} provide further details on the $ \mathcal{DP} $ model and its consequences on Bayesian causal inference. Incorporating these prior assumptions allows for the expected posterior experimental world utility to be computed as:
\begin{equation}
	E_\mathcal{E}[U(B^*,G,\beta)|\bar{b}]= E_\pi[E_\mathcal{E}[U(B^*,G,\beta)|\bar{b}, \pi]] = E_\pi\left[ \frac{1}{C_G}\sum_{i=1}^{n} \sum_{\psi\in \mathcal{I}} \pi_i w_{i}^{\psi}U(b_i,g^\psi,\beta)\right].
	\label{analyticex}
\end{equation}
Note that the right-most expression depends only on observed data, hence the $ ^* $ notation is dropped; this includes dropping the $ ^* $ from the weight $ w^\psi_i $, as it is no longer a random variable but rather an instantiation of that random variable. With this expression for the posterior expected utility, focus turns to maximization. The maximizer of the experimental world expected posterior utility can be obtained by solving: $\beta_{opt}=\arg\max_{\beta}\ E_{\pi} \left[ \sum_{i=1}^{n}\pi_i \sum_{\psi \in \mathcal{I}}  w_{i}^{\psi} U(b_i,g^\psi,\beta)\right]$.

Uncertainty in $\beta_{opt}$  may be characterized by noting that  $\beta_{opt}$ is a deterministic function of  $\pi$, in arguments similar to those in \cite{Walker2010a}. Thus, draws from the posterior distribution of $ \beta_{opt} $ can be done via:
\begin{equation*}
	\beta_{opt}^{*}(\pi)=\arg\max_{\beta}\  \sum_{i=1}^{n}  \pi_i \sum_{\psi \in \mathcal{I}}  w_{i}^{\psi} U(b_i, g^\psi,\beta).
\end{equation*}
This relationship emphasizes that uncertainty in the posterior distribution reflects uncertainty in $\beta_{opt}$. $C_G$ may be disregarded for the purposes of predictive inference. This is an exact Bayesian procedure, modulo Monte Carlo error. Under the specified negative squared error loss utility inference is arrived at by solving:
\begin{equation}
	\beta_{opt}^{*}(\pi)=\arg\max_\beta \left[-\sum_{i=1}^{n} \pi_i  \sum_{\psi\in \mathcal{I}} w_i^{\psi} (y_i-h(\beta,\psi))^2\right].
	\label{eq:solve}
\end{equation}
Equation \eqref{eq:solve} leads to solving for parameters in a similar fashion to how estimating equations are solved in frequentist semi-parametric methods; consequently the $ \mathcal{DP} $ prior with the $ \alpha=0 $ assumption clarifies how solving estimating equations can be interpreted through a Bayesian perspective. Inference about the parameter $ \beta $ in $ h(\beta,\psi) $ requires looking beyond the likelihood times prior formalism that permeates Bayesian inference, though the proposed methods still adhere to the Bayesian inferential principals: a prior is placed on the observational world data-generating mechanism and updated in light of data to obtain a posterior distribution for the observational world data-generating mechanism. Importance sampling and Bayesian decision-theory is then utilized to maximize a expected utility function in the experimental world, and posterior belief about the utility maximizers is propagated from posterior belief about the data-generating mechanism. From equation \eqref{eq:solve}, it is also evident that to draw inference in the experimental world, the weight $w^\psi$ needs to be computed; this leads to modeling the treatment assignment probabilities. For each draw of $ \pi $ a model $p_\mathcal{O}(z_{k}|\bar{z}_{k-1},\bar{x}_{k},\gamma_k(\pi)),\; k=1,...,K$ can be considered. The parameters $\gamma_j$  may be regarded as coming from a posterior utility maximization framework with the same $ \mathcal{DP} $ prior. When the utility is the log-likelihood, the following maximization is required for any $ k=1,...,K $:
\begin{equation*}
	\gamma_{k,opt}^{*}(\pi)=\arg\max_{\gamma_k}\sum_i^n \pi_i \log p_\mathcal{O}(z_{i,k}|\bar{z}_{i,k-1},\bar{x}_{i,k},\gamma_k).
	\label{luli}
\end{equation*}
Note that in this case, the log-likelihood utility does not have a parameter $ g^\psi $. This is because we are now concerned with a utility in the observational world $ \mathcal{O} $. Then, for every draw $\pi$, the weighted treatment propensity model can be fit, and the resulting weight, $w^{\psi}$, in equation \eqref{weight_equation} is now dependent on $ \pi $. {Fitting a model for the treatment (propensity score) requires care, especially when the treatment assignment mechanism is unknown. Austin et al. \cite{austin2015moving} discuss some considerations for estimating treatment propensities, including model fit, variable selection, and diagnostics. Authors indicate that propensity scores should be built with the aim of balancing confounders, and that emphasis should be placed on balancing those confounders thought to have the largest effect on the outcome. Caution should also be taken when including variables that are only predictive of treatment assignment, as this may result in bias and increased variance of estimated treatment effects \cite{myers2011effects}. Effectively, for each draw $ \pi $, computing $  \beta_{opt}(\pi) $ is coupled with computing $  \gamma_{k,opt}(\pi) $. Thus across draws of $ \pi $, uncertainty in $\gamma_j$ is being incorporated into the estimation procedure. From a practical perspective, the \texttt{glm} function in \texttt{R} can be used to fit these models, making use of the \texttt{weights} argument to supply the relevant information; the optimizer for the negative squared error loss utility is the same as that which maximizes the Gaussian likelihood.
There is some flexibility in the specification of $h(\beta,\psi)$. One example is $h(\beta,\psi)=\beta_0+\beta_1\psi + \beta_2\psi^2$, which can be maximized analytically to identify an optimal DTR. To fit this model requires plugging in $h(\beta,\psi)$ into equation \eqref{eq:solve} and solving it. To solve equation (\ref{eq:solve}) requires a data augmentation procedure that duplicates patient data rows for as many regimes as to which they are adherent. This procedure is detailed in \cite{Cain2010} with further considerations for the specification of these models found in \cite{RodriguezDuque2022}. A description for the estimation procedure can be found in Algorithm 1.

The estimation procedure outlined in Algorithm 1 should be used with care. Before this procedure can be implemented, the family of DTRs must be chosen --- a challenging process that should be guided by subject matter expertise to allow for decision rules that are clinically relevant. Additionally, not all clinically relevant families of decision rules can be explored with a given dataset; it is required that a sufficiently large and representative group of patients in the data follow the regimes of interest. This relates to the positivity condition previously discussed. These considerations emphasize the challenge in defining a research question that is meaningful and feasibly explored with the data.

In the remainder of this section, we introduce three additional Bayesian methods of estimation and inference for optimal DTRs (via inverse weighting with a grid search in section 2.2, via a doubly robust grid search approach in 2.3, and using Gaussian processes to emulate the value function in 2.4, with additional considerations for individualized inference in section 2.5 and normalization of IPW weights relevant to the methods of sections 2.2-2.3 in section 2.6). As with dynamic MSMs, the target of inference of the methods in sections 2.2 and 2.3 is a marginal mean, namely the expected outcome under adherence to a regime $ g^\psi $. Importantly, one difference in the terminology used in this paper is that dynamic MSMs allow for parametric models of the value given $ g^\psi $ for a continuum of $ \psi $s, whereas methods in sections 2.2 and 2.3 target the value of a regime, one regime at a time.

\begin{algorithm2e}[H]
	\SetAlgoLined
	\DontPrintSemicolon
	\KwData{{\small$DATA_\mathcal{O}$}  \tcp*[r]{One row per patient; $ n $  patients}}
	\For(\tcp*[f]{Create {\small$AUGDATA_\mathcal{O}$} based on regime adherence}){$\psi \in \mathcal{I}$}{
		Duplicate rows of {\small$DATA_\mathcal{O}$} for patients adherent to regime $g^\psi$\;
		Add column specifying regime index $ \psi $\;
	}
	Posit model for $h(\beta,\psi)$\;	
	\For(\tcp*[f]{B is number of posterior draws}){$i\leftarrow 1$ \KwTo $B$}{
		Draw $\pi=(\pi_1,...,\pi_n)$ from $ Dir(1,...,1)$\;
		Using weighted logistic regression, estimate $p_\mathcal{O}(z_k|\bar{z}_{k-1},\bar{x}_k, \gamma_k, \pi)  \; \forall k$ \;
		Compute weights $w_i^\psi,\; i=1,...,n$, using probabilities in the previous step\;
		Add weights to {\small$AUGDATA_\mathcal{O}$}\;
		Perform regression with mean $h(\beta,\psi)$ and with weights $\pi_i w_i^\psi$ to obtain $ \beta^{*}_{opt}(\pi) $\;
		Maximize $h(\beta(\pi),\psi)$ to obtain a sample from $ \psi_{opt}^{*}(\pi) $\;
	}
	\KwOut{Posterior distribution of $ \psi^*_{opt}$}
	{\small$DATA_\mathcal{O}$} is an input dataset with one row per patient and is used to fit treatment models. {\small$AUGDATA_\mathcal{O}$} is an augmented dataset, where patients are duplicated for each DTR to which they adhere. This dataset is used to run regression for $h(\beta,\psi)$.
	\NoCaptionOfAlgo
	\caption{Algorithm 1: Fitting procedure for Bayesian dynamic MSM and for identifying $ \psi_{opt} $.}
\end{algorithm2e}
\vspace{0.5cm}

\subsection{Optimal DTRs via Bayesian IPW Inference and a Grid Search}

It may be that we want to avoid using the methods in the previous section, as we do not want to model the value function directly with $ h(\beta,\psi) $. This can be because an incorrectly specified model may lead to incorrectly identifying the optimal DTR. One way to avoid this is to use an estimator for the value of a DTR and to perform a grid search for the optimum over the indices $ \mathcal{I} $ in a family. This requires estimating the value under adherence to a regime for a discrete grid of indices, $ \mathcal{I}_{grid} \subseteq \mathcal{I} $. One way to estimate the value of each regime in the grid is to use the Bayesian IPW which uses a similar framework as in the previous section, with a few differences. In particular, posterior predictive inference is paired with IPW to yield an inferential procedure that uses weighting to create an importance sampling projection of $ \mathcal{O} $ into a regime-enforced world were everyone in the study population follows a fixed regime $ g^{\psi} $. This allows us to use data from $ \mathcal{O} $, where not all patients follow the regime of interest, to infer about a regime-enforced world. This regime-enforced world contrasts the previously considered experimental world where patients are randomized to DTRs in a family at baseline. If we use this method of estimation to compute $E_{g^{\psi}}[Y^*| \bar{b}]$ for each regime in the grid, we can then identify the regime yielding the highest value.

As with the Bayesian MSM, an importance sampling argument and a $ \mathcal{DP} $ prior on the observational world data-generating distribution, leads us to the following:
\begin{equation}
E_{g^{\psi}}[Y^*| \bar{b}] =E_\pi[E_{g^{\psi}}[Y^*| \bar{b},\pi]]=E_\pi\left[\sum_{i=1}^{n}  \pi_i w_{i}^{\psi}y_i\right].
	\label{I}
\end{equation}
The weights $ w_{i}^{\psi} $ are computed as in the previous section. Over repeated draws of $ \pi $, we can compute an estimate for $ E_{g^{\psi}}[Y^*| \bar{b}] $ and its associated variability, relying again on the Bayesian bootstrap to provide an appropriate posterior predictive distribution. Defining $ \tilde{y}^{\psi}(\pi)=E_{g^{\psi}}[Y^*| \bar{b},\pi] $, for conciseness, the optimal regime and its associated variability can be obtained by computing $ \psi_{opt}(\pi)=\arg\max_{\psi \in \mathcal{I}_{grid}}\{y^{\psi_1}(\pi),...,y^{\psi_p}(\pi)\} $, where $ p=|\mathcal{I}_{grid}| $, for each draw of $ \pi $. The treatment models can be incorporated into the estimation procedure in the same way as in the previous section.

In practice, for each draw of $ \pi $, treatment models are fit using the entire observed data and the probability that patient $ i $received the treatment they were observed to receive, $p_{ik}(\pi)=p_\mathcal{O}(z_{i,k}|\bar{z}_{i,k-1},\bar{x}_{i,k},\gamma^*_{k,opt}(\pi)) $,  is computed for each decision point. Then, for each regime in $ \mathcal{I}_{grid} $, weights $ w^\psi $ are computed. Patients who do not follow regime $ g^\psi $ will have a weight of zero, meaning they do not contribute directly to the IPW expression. Patients who do follow regime $ g^\psi $ have weights that depend on $ p_{ik}(\pi), \; k=1,...,K $. Although patients who do not adhere to regime $ g^\psi $ do not contribute directly to the IPW expression, they do contribute to the analysis as they inform the treatment assignment models. Once the value of each regime in $ \mathcal{I}_{grid} $ has been estimated, the regime that optimizes the value can be identified in order to identify $ \psi_{opt}^*(\pi)$. This procedure can be repeated over draws of $ \pi $ in order to obtain the posterior distribution of the optimal regime.

\subsection{Optimal DTRs via Bayesian Doubly Robust Inference and a Grid Search}

Another related approach, which has been explored by \cite{RodriguezDuque2022} to identify optimal DTRs, is to perform a grid search using Bayesian posterior predictive inference and the doubly robust (DR)  estimator proposed by \cite{Orellana2010}. Bayesian predictive inference was first paired with doubly robust estimators by \cite{Saarela2016} to estimate the effect of static treatment regimes. Attention is first given to the characteristics of this estimation approach in order to arrive at a Bayesian estimate of the expected outcome under adherence to a regime $ g^{\psi} $. In the context of identifying an optimal DTR, the doubly robust estimator can then be used to estimate the value of a discrete set of regimes in a family indexed by $ \mathcal{I} $ and the optimal regime in the family identified via a grid search, as presented in the previous section. This means that, like in the previous section, a model $ h(\beta,\psi) $ does not need to be specified. In particular, the DR estimator used yields consistent inference when either a set of treatment models is correctly specified or when a set of outcome models is correctly specified.  Thus, in addition to fitting a sequence of treatments models, as is needed with the IPW estimator, the doubly robust estimator requires that a sequence of conditional outcomes  $\phi^{\psi*}_{k}$, $k=1,...,K$ be estimated. These are defined as
\begin{align*}
		&\phi^{\psi *}_{K}(\bar{x}_{K})=E_\mathcal{O}[Y^*|\bar{X}^*_{K}=\bar{x}_{K},\bar{Z}^*_{K}=\bar{g}^{\psi}_{K}(\bar{x}_{K}), \bar{b}] \;  \text{for} \; k=K \; \text{and as}\\
		&\phi^{\psi *}_{k}(\bar{x}_{k})=E_\mathcal{O}[\phi^{\psi*}_{k+1}(\bar{x}_{k+1})|\bar{X}_{k}^*=\bar{x}_{k},\bar{Z}^*_{k}=\bar{g}^{\psi}_{k}(\bar{x}_{k}), \bar{b}] \;\text{for} \; k=K-1,...,1 .
\end{align*}
Note that these expectations are taken with respect to the probability distribution form the observational world, conditional on subjects who have covariate history $ \bar{x}_{k} $ and who followed the regime $g^{\psi}$ up to time $k$. These $ \phi^{\psi*}_k $ can be interpreted as the posterior expected outcome conditional on covariates $ \bar{x}_k $ and treatments $ \bar{z}_k=\bar{g}_k(\bar{x}_k) $ in a world where regime $ g^\psi $ is followed from stage $ k+1 $ to $ K $. We use the $ ^* $ notation on the $ \phi $s to emphasize that they are expectations taken with respect to a posterior distribution. Further details on these quantities can be found in \cite{Orellana2010}. It can be shown via a conditional expectation argument that $ E_{g^\psi}[Y^*|\bar{b}]=E_\mathcal{O}[\phi^{\psi*}_1(X^*_1)|\bar{b}] $, the estimand of interest.

The next section describes how models for $\phi_k^{\psi *}$ may be fit using regression by parameterizing them with $\tau$ such that $\phi^{\psi *}_{k}(\bar{x}_k)=\phi^{\psi *}_{k}(\bar{x}_k;\tau)$. With these models fit, uncertainty in the parameters can be treated analogously to how uncertainty in $\gamma$ is treated: it is made dependent on $\pi$ via the Bayesian bootstrap. Rather than positing a likelihood model as was done for the treatment assignment mechanism, a negative squared error loss utility can be maximized instead. The result is that for every draw of $\pi$, $\phi^*_{k}(\bar{x}_k,\tau(\pi))$ can be estimated.

Now that these outcome models have been specified, it remains to provide an expression that exhibits the double robustness property when the expectation is taken with respect to the true data generating mechanism. Such an expression is obtained from the following equality:
\begin{equation}
	E_{g^{\psi}}[Y^*| \bar{b}]=E_\mathcal{O}\left[\phi^{\psi *}_1(X_1^*) + \sum_{k=2}^{K} w_{k-1}^{\psi*}(\phi^{\psi *}_{k}(\bar{X}^*_{k})-\phi^{\psi *}_{k-1}(\bar{X}_{k-1}^*)) +w^{\psi *}_K(Y^*-\phi^{\psi *}_{K}(\bar{X}^*_{K})) \middle |\; \bar{b} \right].
	\label{form1}
\end{equation}
Then, the expression inside the expectation on the right hand side exhibits the double robustness property if the outcome models $ \phi^{\psi*} $ are correctly specified or if the treatment models in $ w_k^{\psi*} $ are correctly specified. Note that parameters $ \gamma $ and $ \tau $ in the models have been suppressed for brevity. We note that for this expression to possess the desired property, the positivity condition and the no unmeasured confounders assumption in \cite{Orellana2010} must be met. To incorporate the sampling scheme, a single sample from the posterior distribution of the estimand of interest can be obtained by conditioning on a single draw $ \pi $ in order to obtain:
\begin{equation}
	E_{g^{\psi}}[Y^*| \bar{b}, \pi]=\sum_{i=1}^n \pi_i \left[\phi^{\psi*}_{i1}(x_{i1}) + \sum_{k=2}^{K} w_{ik-1}^{\psi}(\phi^{\psi*}_{ik}(\bar{x}_{ik})-\phi^{\psi*}_{ik-1}(\bar{x}_{ik-1})) +w_{iK}^\psi(y_i-\phi^{\psi*}_{iK}(\bar{x}_{iK}))  \right].
	\label{form3}
\end{equation}
By resampling Dirichlet weights, $E_{g^{\psi}}[Y^*| \bar{b}]=E_\pi\left[E_{g^\psi}[Y^*| \bar{b}, \pi]\right]$ and its associated uncertainty can be computed.  Models for the $\phi$s and $w$s  are coupled with $\pi$ and may be incorporated into the inferential process as was done with the IPW estimators of the previous two sections. To arrive at an optimal regime, this DR estimator can be used to perform a grid search for the optimum.

\subsubsection{Fitting Outcome Models}

To fit the outcome models, some additional definitions are required. First, define the function
\begin{align*}
&Q_K^{\psi*}(\bar{x}_K,\bar{z}_K)= E_\mathcal{O}[Y^*|\bar{X}^*_K=\bar{x}_K,\bar{Z}^*_K=\bar{z}_K, \bar{b}]  \text{ and the stage $ K $ pseudo-outcome as}\\	
&\Delta^{\psi*}_K(\bar{x}_K, \bar{z}_{K-1})=Q_K^{\psi*}(\bar{x}_K,\bar{z}_{K-1}, z_K=g^\psi(\bar{x}_K)).
\end{align*}
This is the expected outcome under observed treatment and covariate values, except for at stage $ K $ where treatment is assigned according to regime $ g^\psi $. For the remaining stages $ k=K-1,...,1 $ define
\begin{align*}
  &Q_{k}^{\psi*}(\bar{x}_{k},\bar{z}_{k})=E_\mathcal{O}[\Delta^{\psi*}_{k+1}| \bar{X}^*_{k}=\bar{x}_{k},\bar{Z}^*_{k}=\bar{z}_{k},\bar{b}], \text{ with stage $ k $ pseudo-outcome}\\
  &\Delta^{\psi*}_{k}(\bar{x}_{k}, \bar{z}_{k-1})=Q_{k}^{\psi*}(\bar{x}_{k},\bar{z}_{k-1},z_{k}=g^\psi(\bar{x}_{k})).
\end{align*}
Then, as elaborated on in \cite{tsiatis2019dynamic}, we can compute the quantities of interest through $ \phi_k^{\psi*}(\bar{x}_k)=Q_{k}^{\psi*}(\bar{x}_k, \bar{g}_k(\bar{x}_k)) $ for $ k=1,...,K $. Of course, in practice $ Q_k^{\psi*} $ and $ \Delta_{k}^{\psi*} $ are unknown, consequently regression models for $ Q_k^{\psi*} $ should be fit and $ \Delta_{k}^{\psi*} $ predicted based on these models. Once all models for $ Q_k^{\psi*} $ have been fit, then $  \phi_k^{\psi*}(\bar{x}_k)  $ can be estimated.  The functions in the \texttt{BayesDTR} package render the estimation of these outcome models straightforward, as all that is required is that users specify the stage-specific models for the pseudo-outcomes (or outcome if at the final stage); the package will perform the required computations in order to arrive at a fit for the $ \phi^{\psi*}_{k} $s. This regression approach is one of several ways to fitting the required outcome models, with \cite{tsiatis2019dynamic} expanding on other methods that can be used.

With these definitions, we now provide a two-stage example of how to obtain estimates for the $ \phi_k^{\psi*} $s. For illustrative purposes, we omit notation pertaining to posterior inference, and then comment on how to incorporate this. The estimation procedure begins by specifying the following two models:
\begin{align}
	 Q_2^\psi(\bar{x}_2,\bar{z}_2) =E[&y|\bar{x}_2,\bar{z}_2]= \beta_{21}x_1+(\beta_{22}+\beta_{23}x_1)z_1 + \beta_{24}x_2+(\beta_{25}+\beta_{23}x_1)z_2,\label{out1}\\
	Q_1^\psi(x_1,z_1)=E[&\Delta_2^\psi|x_1,z_1]=\beta_{11}x_1+(\beta_{12}+\beta_{13})z_1,
	\label{out2}
\end{align}
where  $\Delta_2^\psi =E[Y|\bar{X}_2=\bar{x}_2,Z_1=z_1,Z_2=g_2^\psi(\bar{x}_2)]$. We can use, for example, the \texttt{lm} function in \texttt{R} to fit these models. Note that $\Delta_2^\psi$ is not observed and so it must be predicted using the stage two model.
Once these models have been fit, we may compute the outcomes for the doubly robust estimator by using the data and the estimated models to predict:
\begin{align*}
	&\phi^\psi_2(\bar{x}_2)=Q^\psi_2(\bar{x}_2, \bar{g}^\psi(\bar{x}_{2}))= \beta_{21}x_1+(\beta_{22}+\beta_{23}x_1)g^\psi_1(x_1) +
	 \beta_{24}x_2+(\beta_{25}+\beta_{23}x_2)g_2^\psi(\bar{x}_2),\\
	&\phi^\psi_1(x_1)=Q^\psi_1(x_1,g^\psi(x_1))=\beta_{11}x_1+(\beta_{12}+x_1)g_1^\psi(x_1).
\end{align*}

To incorporate the posterior sampling component, it is necessary to additionally weight by $ \pi $ when fitting models in equations (\ref{out1}) and (\ref{out2}) so that the estimated $ \beta $s are dependent on $ \pi $. This can be done through the \texttt{weights} argument in the \texttt{lm} function.These outcomes may then be used in equation (\ref{form3}) to obtain an estimate of the value under adherence to a DTR $ g^\psi $. Over repeated draws of $ \pi $,  $ E_{g^{\psi}}[Y^*| \bar{b}] $ and its associated uncertainty can be computed. Having computed these estimates of the value for all candidate regimes in $ \mathcal{I}_{grid} $, the value-maximizing regime is selected as optimal.

\subsection{Identifying Optimal DTRs via Gaussian Process Emulation}

As discussed in the preceding sections, there are several value-search approaches to identifying optimal DTRs. The value surface can be modeled directly via a dynamic MSM and consequently maximized or a grid search can be employed in order to identify the optimal regime.  Directly modeling the value surface with a dynamic MSM can yield accurate, interpretable results, but this is only guaranteed when the value surface is correctly specified; for example, incorrectly specifying a quadratic MSM can lead to inadequate inference about optimal regimes if the relationship is not in fact quadratic or poorly approximated by such a function over the range of $ \psi $s considered. A grid search also has limitations in that it may not robustly identify the optimal regime, especially when the estimator used exhibits higher variability in some regions of the decision space than in others or when the value surface is multi-modal. In addition, the grid search is not a particularly efficient approach as it requires many estimator evaluations, which may be computationally burdensome, especially in Bayesian settings where posterior predictive quantities must be computed. An important question that arises from these considerations is whether these limitations can be avoided by alternate methods.

One approach recently explored by \cite{RodriguezDuque2022b} is to make use of computer experiments  to identify optimal DTRs. The term "computer experiment" refers to the idea of sampling function values at strategically chosen points in order to approximate the function, with a limited number of samples. In a DTR context, this involves considering a DTR family, indexed by $ \psi \in \mathcal{I} $, selecting an initial set of design points in $ \mathcal{I} $, and using an estimator for the value of a DTR at these points. With a working model for the value surface, more points can be selected sequentially using a criterion that specifies where an optimum may be. Traditional approaches for computer experiments use regression-based methods to approximate a response surface of interest, like the value surface. However, these approaches have been critiqued, for example, by Huang et al. \cite{Huang2006} who emphasize that regression models are often too simple and unlikely to well-represent complex systems over the entire domain. This critique is analogous to the concerns that arise when using smoothly modeled MSMs to identify optimal DTRs.

Contemporary literature on computer experiments focuses on using  $\mathcal{GP}$s to approximate complex functions and to identify optimizing points \cite{Santner2018}. A $\mathcal{GP}$ is a stochastic process where any finite collection of variables in the process has multivariate Normal distribution. Much of the computer experiments literature has centered around settings in which the function to be maximized is known. However, it should be apparent that this is not the scenario under consideration here in the DTR context. In particular, an analyst wishing to perform a DTR analysis does not have access to direct observations of the value function; they have access only to a noisy, estimated version of it. Guan et al.~\cite{guan2020bayesian}, in estimating optimal dental visiting schedules, also use $ \mathcal{GP} $s to perform optimization. The methodology  developed in \cite{RodriguezDuque2022b}, and to be discussed in the following, places emphasis on understanding how these methods may be accessibly adopted in general analysis problems. Careful attention is given to justifying the use of $ \mathcal{GP} $ regression in computer experiments, an important challenge outlined by \cite{Forrester2006}. Discussion about how to account for sources of uncertainty is also emphasized, in contrast to \cite{guan2020bayesian}. Related work using $ \mathcal{GP} $s as surrogates for the value function can be found in \cite{freeman2022dynamic}. With these nuanced differences in mind, we now examine the inferential problem more closely.

In order to better understand the problem characteristics, some terminology regarding the functional relationships in the problem should be set. The target of inference is the \textit{value surface} which represents the relationship between a DTR $ g^\psi $ idexed by $ \psi $ and its value $E_{g^\psi}[Y] $. As the value surface is not accessible, it must be approximated via the \textit{estimation surface}, a surface that results from point-wise evaluation of an estimator to obtain $\hat{E}_{g^\psi}[Y]$ for varying $ \psi \in \mathcal{I} $. Evaluating the estimation surface on a fine grid is not desirable as not all points on the grid provide the same information about the optimal DTR's location. It would be beneficial to have a sample where each data point provides a high level of information toward identifying the optimizing point. Consequently, the aim is to use a restricted number of points from the estimation surface to produce an \textit{emulation surface} which represents posterior belief about the value surface based on the information gathered from the estimation surface, with the goal of performing fewer evaluations than would be needed for the grid search approaches of sections 2.2 and 2.3. As will be clarified in what follows, this posterior belief will be represented by a  $ \mathcal{GP} $. Certainly, the properties of the methodology to be described in what follows hinge on the properties of the estimator utilized to perform the point-wise evaluations, e.g., choosing a biased estimator will likely yield biased results. Consequently, this methodology relies on the appropriate use of those estimators previously described. For example, the propensity score fitting strategies described in Section 2.1 are still crucial for inference.

Another consideration is that belief about the value surface should emphasize some smoothness, however the estimation surface used to infer about the value surface is not smooth. This is because it is the result of point-wise evaluations of an estimator which utilizes a finite sample to generate an estimate. Recent work concludes that this non-smooth or noisy quality may be heteroskedastic and consequently an inferential approach that accounts for this characteristic may be desirable \cite{RodriguezDuque2022b}. Authors in \cite{RodriguezDuque2022b} examine some methods that allow for optimization via $ \mathcal{GP} $, while accounting for the noise structure. They find that a homoskedastic treatment of the problem yields improved results over a $ \mathcal{GP} $ method that does not account for noise or a grid search, while providing comparable results to an approach that allows for heteroskedastic noise  and that is more computationally intensive. The implementation in the \texttt{BayesDTR} package focuses on a homoskedastic treatment of the noise structure in order to perform optimization; we review this here. The estimation process begins by positing that the estimation surface is a noisy version of the value surface which is denoted by $ f(\psi) $:
\begin{equation}
	\upsilon_i=f(\psi_i)+\epsilon_i \;, \; \epsilon_i \sim N(0,\gamma^2), \; i=1,...,m,
	\label{eq:setup}
\end{equation}
with $ m $ being the number of observed points on the estimation surface. Using a Bayesian nonparametric framework, a  $ \mathcal{GP} $ prior is placed on $ f $; this prior allows for $ f $ to belong to a broad class of continuous functions. Practically, this means that for any $\psi$ ,  $f|\psi$ is $ N(\mu_{0},\mathbbm{K})$ with covariance matrix  $ \mathbbm{K}$ computed via a covariance function $k(\psi_i,\psi_j)$ and parameterized by $\eta_f=(\theta_f, \sigma^2_f) $. $ \theta_f $ is a vector, where entries $ \theta_{fd} $ control the correlation between points in the $ dth $ dimension; $ \sigma^2_f $ scales the correlation function to yield the covariance. Bayesian formulations of this problem have been advocated for by \cite{OHagan1999}, who emphasize that uncertainty in $ f $ is not solely aleatory. For example, in a setting where $ \gamma^2=0 $, $ f $ is a "knowable" function in the sense that it can be evaluated at different values of $ \psi $. However, as it has not been evaluated at all values, there is uncertainty about the function's values in the locations where it has not yet been observed. This uncertainty is not sampling uncertainty arising from the variability in output under a sequence of identical experiments. Prior to continuing, some further notation should be defined, recalling that in this problem the units of observation are now sample points from the estimation surface, not sample points $(\bar{x},\bar{z},y)$ relating to patient information which are fixed at a sample size $ n $. In this problem, data are observed as $\mathcal{D}=\{\psi_i,\upsilon_i\}_{i=1}^m$, and the following vectors are defined $ \psi=(\psi_1,...,\psi_m)^T $, $\upsilon=(\upsilon_1,...,\upsilon_m)^T$ and $ f=(f_1,...,f_m)^T$. Recall that $ \psi_i $ is the regime index for the $ i$th regime (sample point) in the sample and that it could be a vector quantity.

Assuming known hyperparameters, the posterior distribution for the value of a new observation
$ \psi_{m+1} $ is given by:
\begin{align}
	\begin{split}
		f^*_{m+1}|&\psi_{m+1},\eta_f, \gamma^2, \mathcal{D}  \sim  N(\mu_{f^*_{m+1}},\sigma^2_{f^*_{m+1}})\\
		&\mu_{f^*_{m+1}}=\mu_0+\mathbbm{k}^T(\mathbbm{K}+\gamma^2 I_m)^{-1}(\upsilon-\mu_{0f})\\
		&\sigma^2_{f^*_{m+1}}=\mathbbm{k}(\psi_{m+1},\psi_{m+1}) - \mathbbm{k}_{m+1}^T(\mathbbm{K}+\gamma^2I_m)^{-1}\mathbbm{k}_{m+1},
	\end{split}
	\label{eq:regress1}
\end{align}
with $ \mathbbm{k}_{m+1} $ being the covariance vector between observed points $ \psi $ and the new point $ \psi_{m+1} $. The posterior distribution for value of an observation on the noisy estimation surface is given by:
\begin{align}
	\begin{split}
		\upsilon^*_{m+1}|&\psi_{m+1},\eta_f ,\gamma^2, \gamma^2_{m+1}, \mathcal{D}  \sim  N(\mu_{\upsilon^*_{m+1}}, \sigma^2_{\upsilon^*_{m+1}})\\
		&\mu_{\upsilon^*_{m+1}}=\mu_{f^*_{m+1}}\\
		&\sigma^2_{\upsilon^*_{m+1}}=\mathbbm{k}(\psi_{m+1}, \psi_{m+1}) - \mathbbm{k}_{m+1}^T(\mathbbm{K}+S)^{-1}\mathbbm{k}_{m+1} + \gamma^2.
	\end{split}
	\label{eq:regress2}
\end{align}
In an empirical Bayes framework, the posterior predictive distribution is given by \[ p(\upsilon^*_{m+1}|\psi_{m+1}, \mathcal{D})= p(\upsilon^*_{m+1}|\psi_{m+1},\eta_f ,\gamma^2, \mathcal{D}  ),\] meaning the parameters are assumed known even though they must be estimated in practice. These are estimated by maximizing the likelihood $ p(\upsilon|\psi,\eta_f,\gamma^2) $. This maximization is performed in the \texttt{BayesDTR} package, using the concentrated likelihood discussed in \cite{Roustant2012} and \cite{Park2001}. The concentrated likelihood is obtained by plugging-in estimated parameters that have maximum likelihood estimates with analytic expressions.  We clarify this in what follows, but first it must be noted that these likelihoods are not always easy to maximize, even with gradient methods, so it is advisable to perform the maximization with several random starting locations as is made possible with the \texttt{DesignFit} function to be discussed in later sections.

In order to maximize the likelihood, the covariance function must first be specified. Common choices for the covariance functions, which yield smooth sample paths, are the  $Mat\acute{e}rn_{3/2}$ and  $Mat\acute{e}rn_{5/2}$ covariances \cite{williams2006gaussian}. The $Mat\acute{e}rn_{3/2}$ covariance function between two regime indices $ \psi_i, \psi_j $ is given by:
\begin{equation*}
	\mathbbm{k}(\psi_i,\psi_j)=\sigma^2_{f}\prod_{d=1}^{D}\left(1+\frac{\sqrt{3}|\psi_{id}-\psi_{jd}|}{\theta_{fd}}\right)\exp\left(\frac{-\sqrt{3}|\psi_{id}-\psi_{jd}|}{\theta_{fd}}\right),
\end{equation*}
where $ D $ is the dimension of $ \psi $ and $ \psi_{id} $ and $ \theta_{fd} $ are the $ d $th entries in the $ \psi_i $  and $ \theta_f $ vectors, respectively. This product emphasizes the point that different candidate rules in $ \mathcal{G} $ should have the same dimension, and each entry in the index should represent the same rule element. Appendix A shows the formula for the $Mat\acute{e}rn_{5/2}$ covariance.

Although empirical Bayes requires maximizing a likelihood dependent on parameters $\mu_0, \eta_f,\gamma^2 $, the maximization is more efficiently performed by changing the parameterization. We now provide this new parameterization; full details of this parameterization can be found in \cite{Roustant2012}. By defining $ \alpha=\sigma^2_f/(\sigma^2_f+\gamma^2) $ and considering the correlation matrix $ R $ defined by $ \mathbbm{K}=\sigma^2_fR $, $(\mathbbm{K}+\gamma^2 I_m)$ can be re-expressed as $v(\alpha R + (1-\alpha)I_m)$, where $ v=(\sigma^2_f+\gamma^2) $. This re-parameterization results in a likelihood dependent on $\mu_{0f}, \theta_f,v, \alpha $, whereas the likelihood in the original parameterization dependended on $\mu_{0f}, \theta_f, \sigma^2_f,$ and $\gamma^2 $. As there are analytic expressions for the optimal $  \mu_{0f} $ and $ v $, the user only needs to concentrate on the maximization in the $ \theta_f $ and $ \alpha $ directions.

Priors for $ \theta_f $ can be incorporated independently for each dimension $ d $, for example, via a Log-Normal prior distribution with parameters $\mu_d, \sigma^2_d $, which can be used to express belief about the size of $ \theta_{fd} $s and consequently the correlation between points. This prior is independent for each $ \theta_{fd} $ and can be added into the log concentrated likelihood with the following term
\begin{equation}
	\sum_{d=1}^{D} -\frac{(\log(\theta_{fd})-\mu_d)^2}{2\sigma^2_d}- \log(\theta_{fd} \sigma_d \sqrt{2\pi}).
\end{equation}
Maximizing the concentrated log likelihood with the added term above amounts to  maximum a posteriori inference, where parameter estimates are fixed at the maximizers of the posterior distribution. One approach to setting the prior hyperparametrs is to identify what a 10\% change is in the direction of interest. Then, the hyperparameters should be chosen such that the 5th and 95th percentiles of the Log-Normal distribution yield a correlation between 0.05 and 0.95. This posits that in the direction of interest, a unit change of 10\% of the range of values will have function outputs that can either be very different from each other or very similar. This is similar to \cite{Lizotte2008}, who set hyperparameter values that prevent the $\theta_{fd}$s from getting very small or very large, thereby preventing that function's value from being nearly exactly correlated or uncorrelated.

\subsubsection{Sequential Sampling and Stopping Considerations}

Recall the desired experimental setup: an initial set of design points are obtained and an initial model is fit on these data. This model, together with a rule for sampling additional points is used to identify new points that are most informative about the optimization process. The rule used to identify new points to sample is generally termed an infill criterion, and a review of possible criteria for stochastic computer experiments can be found in \cite{Picheny2013}. The focus here lies in using the well-known expected improvement criterion \citep{Jones1998} as the infill criterion. In the deterministic setting, Frazier et al. \cite{Frazier2016} mention that this criterion benefits from a result that states that the true optimum will be  identified as the number of experimental points increases, as shown by \cite{Locatelli1997}; this is not guaranteed in the stochastic setting, as uncertainty remains in already observed points, thus requiring for some adaptations.

One solution for this, proposed by \cite{Forrester2006}, is to use a re-interpolation approach. To perform the re-interpolation, the mean
$\hat{\upsilon}_m=E[\upsilon^*_{m}|\psi_{m} ,\mathcal{D}]$ is computed for each observed data point; this results in a new dataset $\mathcal{D}'=\{\psi_i,\hat{\upsilon}_i\}_{i=1}^m$. A $ \mathcal{GP} $ can then be fit on these new data, assuming there is no noise, $ \epsilon $, in the process. The resulting $ \mathcal{GP} $ has the property that there is zero uncertainty at already sampled points, thereby allowing for the use of the expected improvement criterion as a basis for sequential sampling. When the objective is maximization, this criterion is given by  $EI(\psi)=E\left[\max(\upsilon(\psi)-\upsilon_{max})^{+}|\mathcal{D}'\right]$, with  $ \upsilon_{max}=max(\upsilon_1,...,\upsilon_m) $. It is important to understand what this criterion means, in order to understand why it should be maximized to identify new points to add to the sample. At a point $ \psi^{new} $ where $\upsilon_{max}$ is expected to be greater than $ \upsilon(\psi^{new}) $, this criterion is zero. At a point $ \psi^{new} $ where $ \upsilon(\psi^{new}) $ is expected to be greater than $\upsilon_{max}$, this criterion is large, with magnitude increasing with the difference in values. Therefore, maximizing this criterion adds points to the sample that are believed to have a higher value than the currently observed maximizer. Importantly, the expectation is taken with respect to the posterior distribution and it can be further developed to yield the well-known formula:
\begin{equation} EI(\psi)=(\mu_{\upsilon^*_{m+1}}(\psi)-\upsilon_{max})\Phi\left(\frac{\mu_{\upsilon^*_{m+1}}(\psi)-\upsilon_{max}}{\sigma_{\upsilon^*_{m+1}}(\psi)}\right) + \sigma_{\upsilon^*_{m+1}}(\psi) \dot\Phi\left(\frac{\mu_{\upsilon^*_{m+1}}(\psi)-\upsilon_{max}}{\sigma_{\upsilon^*_{m+1}}(\psi)}\right)
\end{equation}
when $\sigma_{\upsilon_{m+1}}(\psi)>0 $ and $ 0 $ otherwise. $\Phi$ is the CDF of the Standard Normal distribution and $\dot\Phi$ is the pdf.

As the expected improvement is zero at each visited point, it is clear that the function exhibits multi-modality. Maximization of this function can be performed via a genetic algorithm, as is done in \cite{Roustant2012}, and implemented by \cite{Walter2011} with the \texttt{rgenoud} package in \texttt{R}.

Finally, one natural question that arises is when to stop sampling. One approach may be to stop sampling when the expected improvement at newly sampled points plateaus near zero. Another approach, which we utilize in the illustrative example in section 4, is to plot the newly sampled points in order of sampling, to determine if sampling has converged around a specific region, which may suggest that the algorithm is sampling in a region where it believes the optimum to be.

\subsubsection{Uncertainty Quantification and Fitting Procedure}
One important element that should be addressed is the quantification of uncertainty in the estimated optimal DTR. In a grid search, uncertainty in the optimum can be estimated via the Bayesian bootstrap. However this can be computationally intractable if the estimator employed arises from a posterior distribution with no analytic expression for the mean, as this requires complex computation for each bootstrap sample. Furthermore, bootstrapping the grid search does not quantify uncertainty arising from the grid size selected. Certainly a coarse grid should have a different level of uncertainty about the optimizer than a fine grid, however, it is not clear how to quantify this.

With the $ \mathcal{GP} $ approach presented, a Bayesian bootstrapping scheme can also be used to quantify sampling uncertainty. It can be further combined with the posterior uncertainty which represents uncertainty in the value function after having  sampled $ m $ points from the estimation surface. For example, for each bootstrapped sample, $ N $ sample paths can be obtained from the posterior distribution and the optimum identified for each of these sample paths. Over bootstrapped samples, the resulting distribution of optima is reflective of both uncertainties. In what follows, we will examine how to quantify uncertainty in this manner with the \texttt{BayesDTR} package. This is of course a computationally intensive procedure.

In Algorithm 2, we provide a full description of how to identify optimal DTRs with the discussed $ \mathcal{GP} $ methodology.

\begin{algorithm2e}[h]
	\SetAlgoLined
	\SetNoFillComment
	\DontPrintSemicolon
	\tcc{First obtain point estimates for $ \psi_{opt} $}
	Estimate value, $ \tilde{y}^{\psi} \coloneqq E_{{g^\psi}}[Y]$, at experimental points $ \mathcal{P}=\{\psi_1,...,\psi_m\} $ \;
	Estimate $ \mathcal{GP} $ parameters \;
	Perform re-interpolation as in \cite{Forrester2006}\;
	
	 \SetKwRepeat{Do}{do}{while}
	 \Do{Not converged  \tcp*[f]{Assess convergence as in section 2.4.1}}{
	 Sample new point by solving $ \psi^{new}=\arg\max_\psi\{EI(\psi); \psi \in \mathcal{I}\} $\;
	 Estimate value at $\psi^{new} $\;
	 Add $\psi^{new} $ to experimental points: $ \mathcal{P}=\{\psi^{new}\} \cup\mathcal{P} $ \;
	 Identify $ \psi^{opt}=\arg\max_{\psi} \{\tilde{y}^\psi; \psi \in \mathcal{P}\} $\;
	 Estimate $ \mathcal{GP} $ parameters and perform re-interpolation;
 	 }
	 Set $ m_{+} =|\mathcal{D}|$ \tcp*[r]{Now have point estimate for $ \psi_{opt} $} \;
	\tcc{Now computing variability around optimal thresholds}
	\For(\tcp*[f]{B is number of Bayesian bootstrap draws}){$i\leftarrow 1$ \KwTo $B$}{
		Draw $\pi=(\pi_1,...,\pi_n)$ from $ Dir(1,...,1)$ \;
		\tcc{Estimates, $\tilde{y}^{\psi}(\pi)$, now depend on $ \pi $ as in section 2.2 and 2.3}
		As above, sequentially sample points by maximizing $  EI(\psi)  $ and updating $ \mathcal{GP} $ parameters\;
		Stop sampling when total of $ m_{+} $ experimental points are in $ \mathcal{P} $\;
		Draw $ N $ sample paths from posterior $ \mathcal{GP} $\;
		Compute optimizer for each sampled path\;
		Store vector of length $ N $, containing $ N $ optimizers\;
	}
	\KwOut{Vector of length $ N\cdot B $ containing posterior distribution of $\psi^{opt}$}
	\NoCaptionOfAlgo
	\caption{Algorithm 2: Finding optimal DTRs using $ \mathcal{GP} $ emulation.}
\end{algorithm2e}

\subsection{Individualized Inference}
The Bayesian methods discussed so far permit individualized inference. This is best understood via an example. Consider the regime "treat if $x>\psi$" and suppose that a new patient is observed with covariate value $x^{new}$. Interest lies in deciding whether this patient should receive treatment, based on what is known about the optimal threshold, $\psi_{opt}$. This involves computing $P(x^{new}>\psi^*_{opt}|\bar{b})$ by taking a sample of size $m$ from the posterior distribution of $ \psi^*_{opt} $ and computing $p=(1/m) \sum_\psi \mathbbm{1}(x^{new}>\psi_i^*)$. Given uncertainty in $ \psi_{opt} $, this measure informs a decision maker about the probability that $ x^{new} $ is above the true optimal threshold. Effectively, then, it provides evidence for whether the patient should receive treatment if the optimal regime is to be followed. This approach is relevant to all types of decision rules. We will see in the illustrative example how to implement this individualized inference about the treatment decision.

\subsection{Frequentist and Normalized Estimators}

The Bayesian approaches discussed in sections 2.1-2.3 all have frequentist counterparts. Point estimates for the quantities of interest can be arrived at in a straightforward manner. For the dynamic MSMs in section 2.1, it is necessary that $ \pi_i \text{ for } i=1,...,n $ be removed from equation (\ref{eq:solve}). Solving this new equation will yield the frequentist point estimates. For the IPW method, it is required that the expectation in equation (\ref{I}) be computed to yield $ \sum_{i=1}^{n} \frac{1}{n} w_{i}^{\psi}y_i $, as $ E_\pi[\pi_i]=1/n $. For the doubly robust approach, it is required that the $ \pi_i $ in equation (\ref{form3}) be replaced with $ 1/n $. Treatment models are now fit without any dependence on $ \pi $.

In practice, using estimators with less variability can improve the resulting inference. In the case of the IPW and DR estimators, it is clear that reducing the variability in the weights will reduce the variability in the estimator. This may be achieved via normalized weights. Weight normalization is discussed in \cite{Hernan2020}, and has been explored in \cite{Xiao2010}, as a means of reducing variability in weighted estimators. In this Bayesian setting, there is a contribution to the weights from the importance sampling weights and from the Dirichlet weights.  For each sample of Dirichlet weights $\pi= (\pi_1,...,\pi_n) $, the normalized weights can be defined as:
\begin{equation}
	\bar{w}^{\psi}_{ik}=\dfrac{\dfrac{\pi_i\mathbbm{1}_{\bar{g}^\psi_k(\bar{x}_{ik})}(\bar{z}_{ik}) y_i  }{\prod_{j=1}^{k}p_\mathcal{O}(z_{ij} |\bar{z}_{ij-1},\bar{x}_{ij})}}{\displaystyle\sum_{i=1}^n\dfrac{
			\pi_i\mathbbm{1}_{\bar{g}^\psi_k(\bar{x}_{ik})}(\bar{z}_{ik})  }{\prod_{j=1}^{k}p_\mathcal{O}(z_{ij}|\bar{z}_{ij-1},\bar{x}_{ij})}}, k=1,...,K.
	\label{normalized}
\end{equation}
Taking the expectation in the numerator and the denominator across $ \pi $  yields the normalized weights that could be used in a frequentist analysis as in \cite{Hernan2020}. Replacing $ \pi_i \bar{w}^{\psi}_{ik}$ in the IPW or DR estimator by the weight in equation (\ref{normalized}) yields the normalized estimators.

\section{Implementation}
In this section, we examine the functions in the \texttt{BayesDTR} package that can be used to carry out inference with the methods described previously. We first examine the \texttt{BayesMSM} function, which permits identification of optimal DTRs using Bayesian dynamic MSMs, IPW, and doubly robust estimators. We then focus our attention on the \texttt{DesignFit} and \texttt{SequenceFit} functions which perform estimation using the  $ \mathcal{GP} $ methodology of section 2.4. We also examine the \texttt{FitInfer} function which allows for the quantification of uncertainty in the optimal DTR when using $ \mathcal{GP} $s. The functions in the package do not restrict the number of decision points in the analysis, though care should be taken when considering a large number of decision points; methods that rely on inverse probability weighting can result in extreme weights when taking the product of treatment probabilities across many decision points. Furthermore, functions in this package allow for a continuous outcome to be optimized at the end of the study, with binary treatment options at each decision point. No restrictions are placed on the form of the tailoring and auxiliary covariates. Importantly, there should be no missing values in any of the variables utilized for analysis.

\subsection{Functions to Identify Optimal DTRs using Bayesian Dynamic MSM, IPW, and Doubly Robust Estimators}
The following code provides the syntax required to use the \texttt{BayesMSM} function.  The \texttt{BayesMSM} function has three distinct functionalities: I) to infer about the parameters of a dynamic MSM via IPW, II) to estimate the value of a grid of regimes via IPW, and III) to estimate the value of a discrete set of regimes via the doubly robust estimator.
\begin{verbatim}
#loading BayesDTR package
library(BayesDTR)
#Basic parameters in the BayesMSM function
BayesMSM(PatID,Data,Outcome_Var,Treat_Vars,Treat_M_List,Outcome_M_List,MSM_Model,
	 G_List,Psi,Bayes=TRUE,DR=FALSE,Normalized=FALSE,B=100,Bayes_Seed=1)
\end{verbatim}
\texttt{PatID} and \texttt{Data} allow users to supply an analysis dataset and to indicate the patient identifier. Note that the analytic dataset should contain only one row per patient. \texttt{Outcome\_Var} is a character variable specifying the final-stage outcome, and \texttt{Treat\_Vars} is a character vector specifying the stage-specific treatment variables. Treatment variables should be coded as \{0, 1\}. \texttt{Treat\_M\_List} and \texttt{Outcome\_M\_List} are lists containing the formulas for the treatment and outcome models, depending on which estimator is being used. In each list, there should be as many formulas as treatment decision points and they should be ordered chronologically. A formula for the MSM of interest can be supplied via the \texttt{MSM\_Model} parameter, if the aim is to make use of functionality I.

The next set of parameters are those relevant to the family of dynamic regime of interest. The \texttt{G\_List} variable allows the user to define the stage-specific decision rules of interest. The \texttt{Psi} parameter is a matrix specifying the DTR index grid that will be used to create an augmented dataset if using a dynamic MSM or to perform a grid search if directly using an estimator for the value. In \texttt{Psi}, there should be one column per regime index coordinate and it is necessary that the column names match the names in the regime indices provided to \texttt{G\_List}. For example if at stage one, the regime of interest is "treat when \texttt{psi\_1>x}", then \texttt{Psi} should contain a column named \texttt{psi\_1}. The rows of \texttt{Psi} corresponds to a single point in the grid. The function can handle decision rules that involve one of five comparison operators per stage \texttt{==,>,<,>=,<=}. On either side of the comparison operator, there can be parameters that index the family of regimes, or there can be tailoring covariates. The parameters and tailoring covariates can appear in the same expression with the usual mathematical operators for example as given by the rule "treat when \texttt{psi\_1*x\_1+psi\_2*x\_2>0}". Lastly, the \texttt{Bayes}, \texttt{DR}, and \texttt{Normalized} parameters indicate whether a Bayesian or frequentist analysis should be carried out, whether or not the DR estimator should be used, and whether weights should be normalized or not. \texttt{B} allows the user to indicate the number of Bayesian bootstrapped samples to perform when  \texttt{Bayes=TRUE}. The default fit for this function is to use a grid search with the IPW estimator.

The Bayesian analysis returns a matrix containing the posterior distribution of interest. For functionality I, there are as many columns as terms in the \texttt{MSM\_Model} formula, and the number of rows is equal to \texttt{B}. Each matrix column represents a sample from the posterior distribution for a regression coefficient. For functionalities II and III, columns in the matrix represent points in the grid  of \texttt{Psi} and the rows, like in functionality I, represent distinct posterior draws.
The frequentist analysis returns point estimates, with columns representing the same parameters as in the Bayesian analysis. If the user is interested in providing a measure of variability for the frequentist estimates, the nonparametric bootstrap can be used by calling the \texttt{BayesMSM} function within the bootstrapping function in the \texttt{boot} package.
The illustrative example in Section 4 will provide clarity as to the required format that variables should be provided in and how to perform each of the analyses of interest.  Additionally, a systematic description of required and optional parameters for each functionality is provided in Appendix B.

\subsection{Functions to Identify Optimal DTRs using Gaussian Processes Emulation}
We now examine the syntax for functions in the  \texttt{BayesDTR} package used to identify optimal DTRs using $ \mathcal{GP}s $. The first function we examine is the \texttt{DesignFit} function which allows us to fit a $ \mathcal{GP}$ model on an initial set of design points.
\begin{verbatim}
DesignFit(PatID,Data,Outcome_Var,Treat_Vars,Treat_M_List,Outcome_M_List,
	G_List,Psi,Normalized=TRUE,DR=FALSE,
	Numbr_Samp,IthetasU,IthetasL,Covtype,
        Likelihood_Limits,Prior_List=NA, Prior_Der_List=NA)
\end{verbatim}
The parameters used on the first two lines of the \texttt{DesignFit} function above are those already introduced with the \texttt{BayesMSM} function. In particular, these allow the user to utilize the frequentist IPW or DR  estimator to produce the estimation surface. Note that the \texttt{MSM\_Model} parameter should not be used in this application, as the $ \mathcal{GP} $ only makes use of the IPW or DR estimator for the value of a single regime at a time. If a value for this parameter is passed to the function, it will be ignored and the default normalized IPW estimator will be used. Other required parameters used by the function include \texttt{Numbr\_Samp} which specifies the number of random starts when optimizing the Gaussian likelihood and \texttt{Covtype} which specified whether the $Mat\acute{e}rn_{3/2}$ (\texttt{Covtype}=1) or $Mat\acute{e}rn_{5/2}$ (\texttt{Covtype}=2) covariance functions will be used. The user should also provide the limits for the parameter coordinates in $ \theta_{f}$ in the likelihood via \texttt{IthetasL}, \texttt{IthetasU}, where \texttt{L} stands for the lower bound of the parameter and \texttt{U} stands for the upper bound. Placing bounds in the optimization is useful, otherwise the gradient optimizers may explore a region of the parameter space that yields a non-invertible covariance matrix, thereby interrupting the optimization procedure. As we are interested in assessing whether the model is being fit correctly, the \texttt{Likelihood\_Limits} parameter is a list of  vectors allowing the user to set the limits for plotting the likelihood for the $ \theta_{f} $ coordinates. Each list element is a vector containing the lower and upper bound for each covariance parameter.

Independent priors can also be placed on these parameters by defining the optional parameters \texttt{Prior\_List} and \texttt{Prior\_Der\_List}. These are lists containing the formula for the log prior distributions and for the derivatives of the prior. Importantly, a specific naming convention for elements of these lists should be maintained. For example, in a two-stage setting, the $ \theta_{f1} $ parameter should be represented by \texttt{theta1} and the $ \theta_{f2} $ parameter should be represented by \texttt{theta2}. Adding more dimensions to the problem simply requires adding more elements to the list and maintaining the naming convention. Appendix C provides a systematic description of which parameters are required and which are optional. Note that by default, these optimization functions identify a DTR that maximizes the value function. If the objective is to minimize the value function, users should supply the negative of the outcome variable to \texttt{Outcome\_Var} and allow the function to maximize the value.

The \texttt{DesignFit} function returns a list of several important parameters. In particular, it returns an \texttt{Update} list containing information about the updated $ \mathcal{GP} $ fit as well as a \texttt{ReInter} list containing the \texttt{x\_max\_ri} and \texttt{Y\_max\_ri} values corresponding to the optimal regime index and value identified with the currently available experimental points. The function also returns the parameter values related to the estimated hyperparameters, these can be found in the \texttt{thetas} and \texttt{alpha} parameters.

Now we explore how to sequentially sample additional points using the \texttt{SequenceFit} function which identifies new points to sample by maximizing the expected improvement and then re-estimates the $ \mathcal{GP} $ parameters based on the new information.
\begin{verbatim}
SequenceFit(Previous_Fit,Additional_Samp,
	 Control_Genoud=list(Domain=matrix(c(200,200,500,500),ncol=2)))
\end{verbatim}
All parameters in the \texttt{SequenceFit} function are required with the first being the \texttt{Previous\_Fit} parameter which stores an object returned by either the \texttt{DesignFit} function or the \texttt{SequenceFit} function. Being able to supply an object returned by the  \texttt{SequenceFit} function is important as we may want to continue sampling sequentially even after we have called this function once. Effectively, all options in the object passed to the \texttt{Previous\_Fit} parameter are inherited in the \texttt{SequenceFit} function. The \texttt{additional\_samp} parameter allows the user to tell the function how many additional samples to take sequentially. The last parameter in the function is relevant to the optimization of the expected improvement via the genetic algorithm as implemented by the \texttt{genoud} function. This is the \texttt{Control\_Genoud} parameter which is a list of parameters to be passed to the \texttt{genoud} function. Importantly, the only required parameter to be passed to the \texttt{genoud} function is the \texttt{Domain} parameter which carries information about the domain of the regime index set $ \mathcal{I} $ for which optimization will be performed; it should be a matrix with number of columns equal to the dimension of the regime index, and with columns indicating the lower and upper boundary in each dimension. The specific values that this parameter takes will be dictated by the family of regimes investigated, which itself is dependent on the applied question of interest. The \texttt{SequenceFit} function returns the same object as the \texttt{DesignFit} function, with the addition of the \texttt{EI\_hist} parameter. This parameter contains the expected improvement value at each of the sequentially sampled points.

One option that may be of interest to a user is to compute the posterior mean after arriving at a $ \mathcal{GP} $ fit. This can be done via the  \texttt{PostMean} function.
\begin{verbatim}
PostMean(X,GP_Object)
\end{verbatim}
This function only requires that an object returned from the \texttt{DesignFit} or \texttt{SequenceFit} functions be supplied to \texttt{GP\_Object}, in addition to a parameter $ X $ specifying a coordinate at which to evaluate the mean.

Lastly, as discussed in section 2.4.2, it may be important to provide a measure of uncertainty when identifying the optimal DTR. The function \texttt{FitInfer} allows the user to do this.
\begin{verbatim}
FitInfer(Design_Object,Boot_Start,Boot_End,N,Psi_new,Location,Additional_Samp)
\end{verbatim}
This function requires a \texttt{Design\_Object} parameter, which is the object returned by the \texttt{DesignFit} function. The \texttt{Boot\_Start} and \texttt{Boot\_Stop} parameters allow the user to specify the number of Bayesian bootstrapped samples, for example from \texttt{Boot\_Start=1} to \texttt{Boot\_End=100}, all while allowing for reproducibility as each bootstrapped sample is linked to a specific seed for random number generation. If we were interested in reproducing only bootstrap number 50, we could set \texttt{Boot\_Start=50} and \texttt{Boot\_End=50}, and run the function. Furthermore, the \texttt{N} parameter tells the function how many sample paths to obtain from the posterior $ \mathcal{GP} $ at each bootstrapped sample. The \texttt{Additional\_Samp} parameter is the same as that in \texttt{SequenceFit} function, and the \texttt{Psi\_new} parameter is the grid of points for which to search for an optimum in each sampled path drawn from the posterior $ \mathcal{GP} $. It also determines the dimension of the covariance matrix used to generate the sampled paths, so a very fine grid may be computationally intractable. The only optional parameter in this function is the \texttt{Location} parameter which allows the user to specify where the output of the function should be saved.

This function returns a matrix with number of columns equal to the number of regime index elements plus one. The last column corresponding to the optimal value, and the prior columns correspond to the estimated optimal index. The number of rows in the matrix is \texttt{N(Boot\_Start-Boot\_End+1)}, as for each bootstrapped sample there are $ N $ posterior paths sampled and an optimum identified for each path.

As we are dealing with a $ \mathcal{GP} $, which depends on a covariance matrix that needs to be inverted, numerical issues may arise. For example, when using the \texttt{SeqFit} function to sequentially sample points, users should take care to check that the sampling has not focused on a very specific region. If it has, this is evidence for convergence of the algorithm and may lead to a non-invertible covariance matrix if too many points are sampled in the same region. This non-invertability arises as nearby points can exhibit nearly perfect correlation. This issue can be carried downstream to the quantification of uncertainty if convergence for some bootstrapped samples is achieved faster thereby possibly yielding non-invertible matrices. Issues with non-invertibility are mainly numerical; conceivably, given enough precision in the matrix computation, matrices would be invertible.

\section{Illustrative Example with the BayesDTR Package}
For illustrative purposes, we adapt data from Hammer et al. \cite{Hammer1996} to demonstrate how the discussed methods may be applied with the \texttt{BayesDTR} package. These data originate from a double-blinded randomized trial performed to compare treatments using single and double nucleosides as a means of treating HIV type 1. Focus is given to patients' CD4 cell count which provides a measure of the health of patients' immune system, with higher values indicating better health.  Study enrollment required patients to have CD4 cell counts between  200 and 500 cells$/mm^3 $. A total of 2467 patients were assigned to daily doses of one of four treatments 1) 600 $ mg $ of zidovudine, or 2) 600 $ mg $ of zidovudine \& 400 $  mg $ of didanosine, or 3) 600 $ mg $ of zidovudine \& 2.5 $ mg $ zalcitabine, or 4) 400 $ mg $ didanosine. Variables found in the dataset include patients' race, sex, baseline CD4, 20 week CD4, weight, age, history of antiretroviral therapy, symptoms of HIV infection, and Karnofsky score. These data may be accessed via the \texttt{LongCART} package in \texttt{R} \citep{Kundu2021}.

We restrict our analysis to the use of two dual-therapies, in order to determine which patients should be given zidovudine with zalcitabine versus zidovudine with didanosine, coded as 1 and 0, respectively. It is possible to analyze the data with all four levels of treatment, and scientifically less relevant, as mono-therapy has not been the standard of care for over two decades. Further, the package does not currently support multinomial regression to model multiple levels of treatment, although the conceptual effort to extend to this case is not too great. In particular, we examine whether tailoring therapy on baseline and 20 week CD4 cell counts yields improved 90 week CD4. As the original trial involved treatment assignment only once, we perform a plasmode simulation that randomly assigns an additional treatment decision point at 20 weeks. Variables for this analysis did not exhibit any missing values. There were 524 patients in the zidovudine \& zalcitabine arm and 522 in the zidovudine \& didanozine arm. The known stage-specific treatment probability was 0.5 by design, however we estimate these probabilities, as this can improve efficiency when using IPW estimators \cite{Henmi2004}. As we added an additional treatment variable, we also simulate the final outcome which depends on $ cd4.0 $ and $ cd4.20 $ variables representing baseline and 20 week CD4 cell count, respectively, a $ sex $ variable that equals 1 for males and 0 for females, and treatment variables $ z_1 $ and $ z_2 $. This outcome is deterministically generated by:
\begin{equation}
	y=max(0, 0.2(5cd4.0+ 6sex+ (-3000+9cd4.0)z_1+(-3000+9cd4.20)z_2)).
	\label{outcomeeq}
\end{equation}
For illustrative purposes, we allow $ y $ to represent the final outcome which we take to be 90 day CD4 cell count and the aim is to maximize this value. Without the $ max() $ function in this data generating mechanism, a small proportion of values would be negative, which is not meaningful given that the outcomes represent a cell count. These adapted data can be found in the  \texttt{BayesDTR} package via the \texttt{BayesDat} dataset. The specific regime that we explore is "at each stage, assign to zidovudine with zalcitabine if CD4 cell count is greater than $ \psi_k, \text{ for } k=1,2 $". $ \psi_1 $ and $ \psi_2 $ are restricted to vary between 200 and 500 $cells/mm^3 $. A regime like this may be of interest if a patient requires one therapy when CD4 cell counts are low and another therapy when CD4 cell counts are closer to stable levels. As we know the data-generating mechanism given in equation (\ref{outcomeeq}), we can compute the mean outcome under adherence to a specific regime $ g $ by setting $ z_1=g_1(x_1) $ and $ z_2=g_2(x_2) $ and plugging these values into the equation. Doing so for a fine grid of regime indices allows us to produce Figure \ref{Value_Function}, an approximation for the value function. Employing a grid search with a grid of increments of five yields an optimal regime at $ (\psi_1,\psi_2)=(335,335) $ with an optimal value of 610 cells$/mm^3  $. This matches very closely to the regime obtained theoretically, if we assume that the effect of truncating a small set of negative values at zero is small. With this assumption, the theoretical optimum lies at $ (333.3,333.3) $. In what follows, we will compare this optimum, which is a very good approximation for the true optimal regime, with optima estimated via other methods.
\begin{figure}[H]
	\centering
		\includegraphics[scale=0.2]{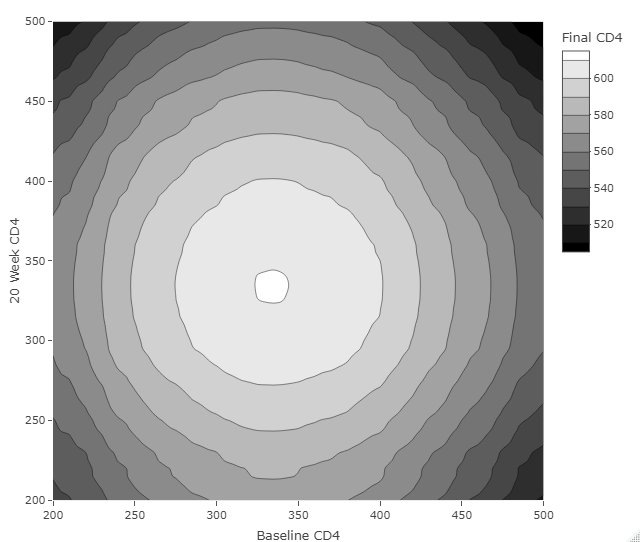}
		\caption{Value function for the rule "treat with zidovudine and zalcitabine when CD4 cell count is greater than $ \psi_k $ for $ k=1,2 $" found via Monte Carlo methods using the known data generating mechanism.}
	\label{Value_Function}
\end{figure}

\subsection{Bayesian MSM, IPW, and Doubly Robust Inference}
We now examine how to define the parameters in the \texttt{BayesMSM} function in order to analyze these data using each of the three estimation approaches available in the function. In the treatment models, we include variables that may not have achieved balance by chance, even though in these data treatment was randomized. In the outcome models, we include these variables as well and additionally include the variables that interact with treatment and that therefore allow for tailoring. Of course, the outcome models are very slightly misspecified, due to the truncation at zero in the data-generating mechanism. However, we will see that this does not appear to have a serious impact on the results.

\begin{verbatim}
#identifying variables in dataset
Outcome_Var="cd4.outcome"; Treat_Vars=c("z1","z2"); PatID="pidnum"
#defining treatment models
Treat_M_List=list(tformula1="z1~karnof+race+gender+symptom+str2+cd4.0+wtkg",
		  tformula2="z2~karnof+race+gender+symptom+str2+cd4.20+wtkg+z1")
#defining outcome models
Outcome_M_List=list(
    oformula1="Pseudo_Outcome~karnof+race+gender+symptom+str2+cd4.0+z1+cd4.0:z1",
    oformula2="cd4.outcome~karnof+race+gender+symptom+str2+cd4.0+cd4.20+z1+
   			cd4.0:z1+z2+cd4.20:z2")
#defining stage specific decision rules
G_List=list(g1=expression(cd4.0>=psi1),
            g2=expression(cd4.20>=psi2))     	
#defining MSM model for when directly modeling the value function
MSM_Model="cd4.outcome~1+psi1+I(psi1^2)+psi2+I(psi2^2)"
\end{verbatim}
For lists defined above, the naming convention of the list elements must be preserved (e.g., treatment models being named \texttt{tformula1}, \texttt{tformula2}, etc.). Furthermore, provided formulas should follow general conventions for formulas supplied to the \texttt{glm} function. Models in \texttt{Treat\_M\_List} are each fit using logistic regression with the \texttt{glm} function and with parameter \texttt{family="binomial"}; models in \texttt{Outcome\_M\_List} are each fit using the \texttt{lm} function as the outcomes are assumed to be continuous. The expressions in the \texttt{G\_List} parameter should contain the conditions for receiving the treatment coded as 1. For functionality I, the target of inference is the coefficients associated with the terms in the model supplied to \texttt{MSM\_Model}. Based on the model supplied above, there are five coefficients of interest, each corresponding to one of the terms supplied to \texttt{MSM\_Model} and here referred to as $\beta_0,...,\beta_4 $. We now provide code for calling the function when estimating optimal DTRs using each of the three estimation methods from section 2.1-2.3. We first examine code for directly modeling the value function by fitting a Bayesian dynamic MSM with IPW.
\begin{verbatim}
#defining grid for augmented dataset
Psi=as.matrix(expand.grid(seq(200,500,50),seq(200,500,50)))
colnames(Psi)=c("psi1","psi2")
#fitting quadratic MSM
QuadMSM=BayesMSM(Data=BayesDat,PatID=PatID,Outcome_Var=Outcome_Var,
	 Treat_Vars=Treat_Vars,Treat_M_List=Treat_M_List,
	 G_List=G_List,Psi=Psi,MSM_Model=MSM_Model,Bayes=TRUE,B=100)
\end{verbatim}
The code above defines the grid upon which to create the augmented dataset required to fit a dynamic MSM, as outlined by \cite{Cain2010}. Here, \texttt{psi1} and \texttt{psi2} index the family of regimes of interest and they match with the variable names defined in \texttt{G\_List}. The column names of \texttt{Psi}, representing coordinates of the regime index, must be named exactly as they appear in \texttt{G\_List}; the function checks for this match in labels, and it will produce a warning if the names do not match. This function call also requires supplying the \texttt{MSM\_Model} parameters and setting the \texttt{Bayes} parameter to \texttt{TRUE} to obtain the Bayesian estimator. Additionally, setting \texttt{B=100} returns 100 posterior draws of the parameters associated with the MSM supplied by \texttt{MSM\_Model}. Note that the \texttt{Normalized} parameter cannot be used when the \texttt{MSM\_Model} parameter is supplied. This function call returns a matrix with columns representing $ \beta_0,...,\beta_4$ corresponding to the MSM specified by \texttt{MSM\_Model}. Rows in this matrix correspond to posterior draws of the $\beta$s. To identify the optimal regime, the quadratic function are maximized for each posterior draw; this yields the posterior distribution of the optimum. We now use the \texttt{BayesMSM} function to estimate the value of a grid of DTRs via the IPW or DR estimator. These are the second and third functionalities available in the \texttt{BayesMSM} function.
\begin{verbatim}
#defining grid for grid search
Psi=as.matrix(expand.grid(seq(200,500,15),seq(200,500,15)))
colnames(Psi)=c("psi1","psi2")
#fitting IPW estimator to a discrete set of regimes
Grid_IPW=BayesMSM(Data=BayesDat,PatID=PatID,Outcome_Var=Outcome_Var,
	 Treat_Vars=Treat_Vars,Treat_M_List=Treat_M_List,G_List=G_List,
	 Psi=Psi,Bayes=TRUE,Normalized=TRUE,B=100)
#fitting DR estimator to a discrete set of regimes
Grid_DR=BayesMSM(Data=BayesDat,PatID=PatID,Outcome_Var=Outcome_Var,
	 Treat_Vars=Treat_Vars,Treat_M_List=Treat_M_List,Outcome_M_List=Outcome_M_List,
	 G_List=G_List,Psi=Psi,Bayes=TRUE,Normalized=TRUE,DR=TRUE,B=100)
\end{verbatim}
In the code above, we first define the grid used for the grid search. We then call the \texttt{BayesMSM} function to return the posterior samples using the IPW and DR estimators. The main difference between the two calls is that the DR approach needs the added parameter \texttt{Outcome\_M\_List} and setting \texttt{DR=TRUE}. For both of these estimation procedures, the \texttt{BayesMSM} function returns a matrix where each column represents a regime index in the same order as provided by the \texttt{Psi} parameter and where each row represents a single draw from the posterior distribution. \texttt{Normalized=TRUE} indicates that we are using normalized weights.

Based on the function calls above, we can estimate the mean value for each of the regimes in the grid; the result is given in Figure \ref{Value Surface}. We see that all methods agree about the general shape of the value function and that in this case both the doubly robust and MSM yield relatively smooth, interpretable surfaces. As with any posterior distribution, summary statistics can be provided. To obtain the posterior distribution for the optimal DTR, the regime index that yields the highest value should be identified for each posterior sample. Doing this across all posterior samples yields the posterior for the optimum. This is shown in the following code for the doubly robust analysis:
\begin{verbatim}
#obtaining index of regime that maximizes value for each posterior sample
max_index=apply(Grid_DR,1,FUN=function(X){which(X==max(X))})
#obtaining posterior distribution of value at optimum
max_val=apply(Grid_DR,1,max)
#obtaining posterior distribution for stage 1 and stage 2 optimal thresholds
Psi[max_index,]
\end{verbatim}

Table \ref{TabResults} shows the posterior median and the 95\% credible intervals for the optimal stage-specific threshold using each of the three described methods. We see that broadly all three methods agree regarding the location of the optimal thresholds. The doubly robust estimator is best at identifying the stage two optimal parameter whereas the quadratic MSM is best in identifying the first stage parameter. All methods seem to perform better at identifying the second stage parameter than the first stage parameter. We also see that the optimal DTR estimated by the DR estimator exhibits less variability than the IPW estimator, which is known to possess the most variability.

\begin{figure}[H]
	\centering
	\hspace{-1.3cm}
		\begin{subfigure}[b]{0.3\linewidth}
		\includegraphics[scale=0.24]{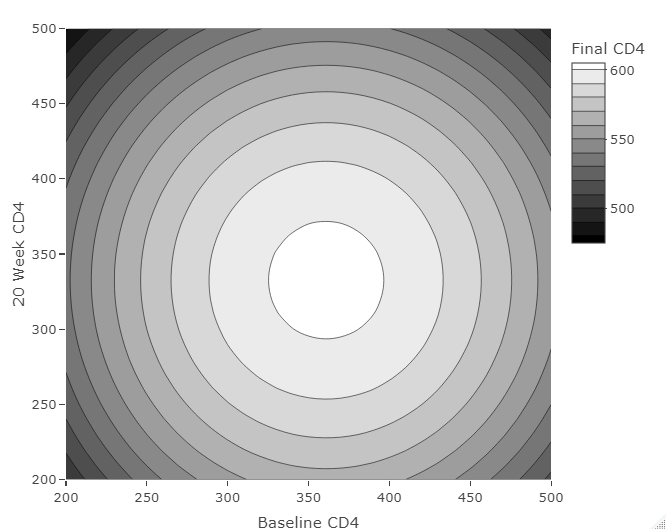}
		\caption{}
	\end{subfigure}
	\begin{subfigure}[b]{0.3\linewidth}
		\includegraphics[scale=0.24]{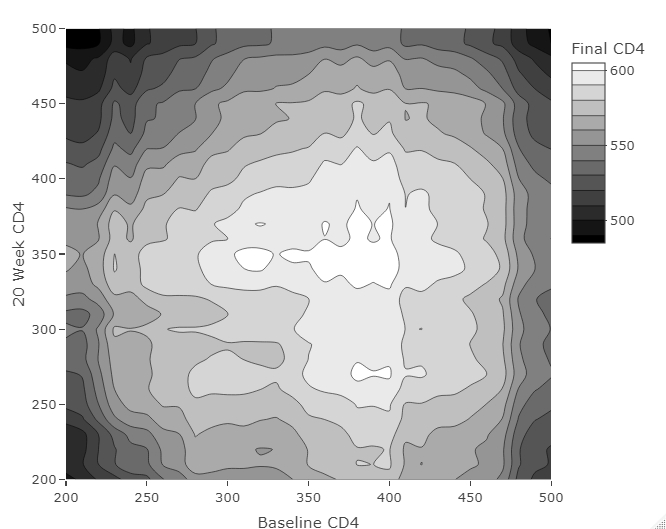}
		\caption{}
	\end{subfigure}
	\begin{subfigure}[b]{0.3\linewidth}
		\includegraphics[scale=0.24]{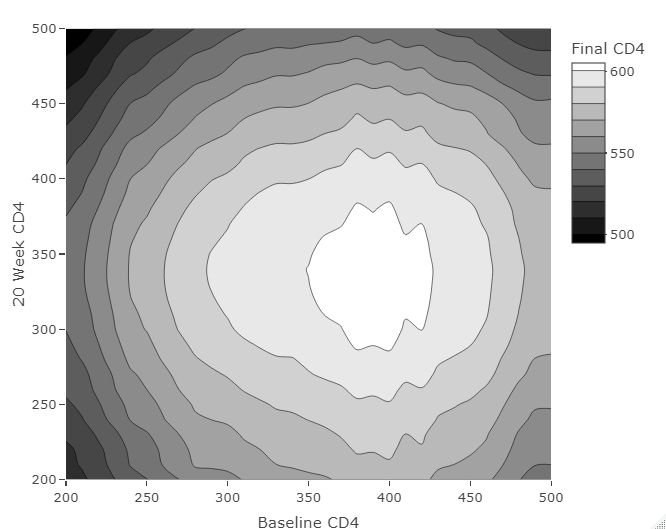}
		\caption{}
	\end{subfigure}
	\caption{Estimation surface for the rule "treat with zidovudine and zalcitabine when CD4 cell count is greater than $ \psi_k $ for $ k=1,2 $" using (a) quadratic MSM  (b) normalized IPW grid search (c) normalized DR grid search.}
	\label{Value Surface}
\end{figure}

\begin{table}[H]
	\centering
	\caption{Estimated optimal thresholds with 95\% credible intervals for the rule "treat with zidovudine and zalcitabine when CD4 cell count is greater than $ \psi_k $ for $ k=1,2 $".}
	\begin{tabular}{llll}
		\hline
		Method & $ \hat{\psi}_{1opt} $ & $ \hat{\psi}_{2opt} $ & Value at Optimum \\
		\hline
		Quadratic MSM     			 & 361 (337,390)         & 332 (306,358)        & 604  (569,643)   \\
		Normalized IPW grid search   & 380 (250,400)  	     & 350 (260,400)       	& 613  (579,652)   \\
		Normalized DR grid search    & 380 (330,420)         & 340 (330,340) 		& 608  (582,633)  \\
		$ \mathcal{GP} $			 & 380 (335,410)         & 335 (245,395)	 	& 607  (569,646) \\
		\hline
	\end{tabular}
	\label{TabResults}
\end{table}

Individualized inference can also be implemented. The code below illustrates how this is done for the first stage. First, the posterior distribution for $ \psi_{1opt} $ is computed, in this case we use the \texttt{Grid\_DR} matrix returned from the \texttt{BayesMSM} function. Then, the probability that a patient's baseline CD4 cell count is greater than the optimal threshold is obtained. Below, we compute this probability for a range of CD4 values, \texttt{Psi1\_Grid},that correspond to new patients in order to obtain the \texttt{probas1} variable.
\begin{verbatim}
#computing posterior distribution for optimal index
max_index=apply(Grid_DR,1,FUN=function(X){which(X==max(X))})
#defining range of new cd4 cell counts
#this range is not for one patient but for a set of new patients with varied CD4 counts
Psi1_Grid=seq(200,500,5)
#computing the probability that a patient's baseline CD4 cell count is greater
#than the optimal stage 1 threshold
probas1=sapply(Psi1_Grid, FUN=function(X, max_index){mean(X>Psi[max_index,1])},
        max_index=max_index)
\end{verbatim}
Having computed these probabilities, we can plot them to better visualize the uncertainty. Figure \ref{Individualized_Rule} shows the first stage probabilities associated with each of the estimation methods discussed and for a range of CD4 cell values that can correspond to newly seen patients. We see that the plot associated with the quadratic MSM is smoothest and displays the most certainty about the optimal treatment allocation, as evidenced by the narrow window of the threshold over which the probabilities are farther from 0 or 1.
\begin{figure}[H]
	\centering
	\hspace{-1.3cm}
	\begin{subfigure}[b]{0.35\linewidth}
		\includegraphics[scale=0.35]{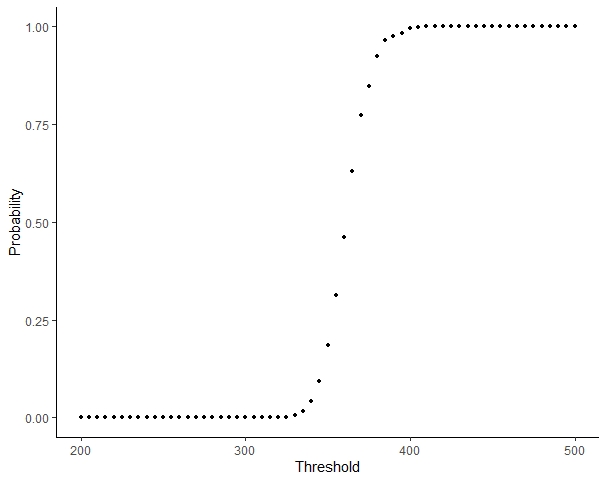}
		\caption{}
	\end{subfigure}
	\begin{subfigure}[b]{0.35\linewidth}
		\includegraphics[scale=0.35]{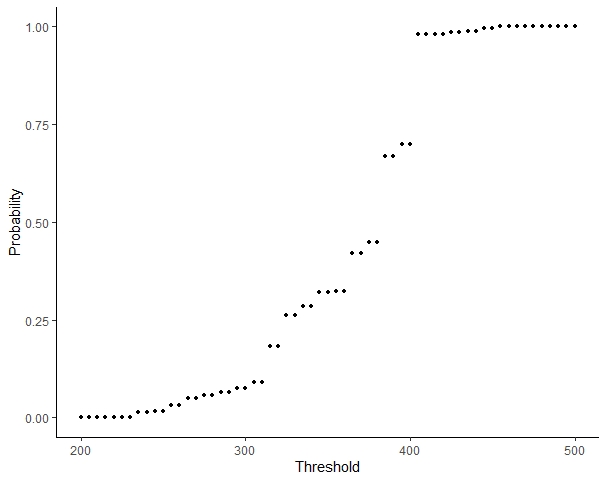}
		\caption{}
	\end{subfigure}
	\begin{subfigure}[b]{0.35\linewidth}
		\includegraphics[scale=0.35]{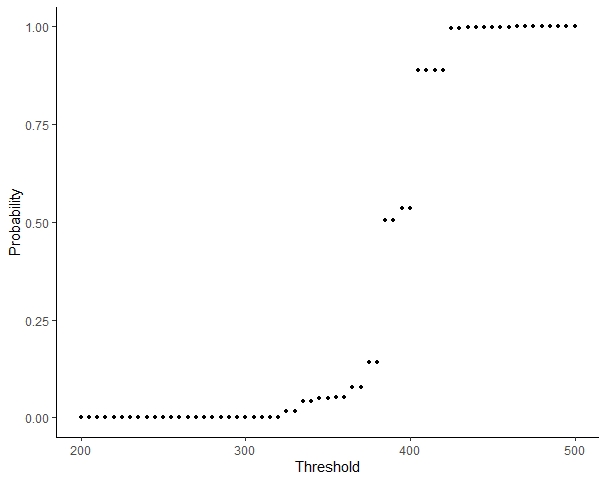}
		\caption{}
	\end{subfigure}
	\caption{Individualized optimal treatment allocation probabilities for rule "treat with zidovudine and zalcitabine when CD4 cell count is greater than $ \psi_k $, $ k=1,2 $" (a) quadratic MSM (b) normalized IPW grid search  (c) normalized DR grid search. }
	\label{Individualized_Rule}
\end{figure}

\subsection{Illustrative Example using Gaussian Processes}
In this section, we continue using the case study presented to examine how to use the \texttt{BayesDTR} package to implement an analysis that uses $ \mathcal{GP} $ emulation to identify an optimal DTR. We begin by fitting a $\mathcal{GP}$ model on an initial set of design points; the required code is given below. The initial set of design points in this setting is in \texttt{Psi} and is limited to 16.

\begin{verbatim}
#creating grid of initial design points
Psi=as.matrix(expand.grid(seq(200,500,100),seq(200,500,100)))
colnames(Psi)=c("psi1","psi2")

#fitting GP model on initial set of design points
start_fit=DesignFit(PatID=PatID,Data=BayesDat,Treat_M_List=Treat_M_List,
	 Outcome_Var=Outcome_Var,Treat_Vars=Treat_Vars,G_List=G_List,Psi=Psi,
	 Numbr_Samp=5,IthetasU=c(600,600),IthetasL=c(0.01,0.01),Covtype=2,
	 Likelihood_Limits=list(seq(250,500,2), seq(250,500,2),
	 Prior_List=NA, Prior_Der_List=NA))
\end{verbatim}

With a $ \mathcal{GP} $ process model being fit on an initial set of design points, the next step is to sample an additional set of experimental points by maximizing the expected improvement. This is done with the \texttt{SequenceFit} function. In this case we select an additional six points.
\begin{verbatim}
#Updating model with newly sequentially sampled points
second_fit=SequenceFit(Previous_Fit=start_fit,Additional_Samp=6,
       Control_Genoud=list(Domain=matrix(c(200,200,500,500),ncol=2)))
\end{verbatim}
Once additional samples have been obtained, the sample points can be plotted to examine whether the algorithm has focused sampling in a specific region, thereby providing evidence that a maximizer has been identified. Figure \ref{Convergence_Assessment} shows these plots. The first 16 points correspond to the design points, the remaining six points correspond to those sequentially sampled; we see that the sequentially sampled points have remained very much in the same area thereby providing evidence for convergence. Note that it does not matter what order the first 16 points are plotted in, as they were sampled simultaneously. Users should be cautious about how many additional points to sample in settings like this, as sampling points that are proximal to each other can result in non-invertible covariance matrices. Using the model fit of six additional points, we can determine the optimal thresholds and the value at the optimal thresholds. \texttt{Y\_max\_ri} gives the value at the optimum to be 601.8 cells$/mm^3  $; the optimizer can be found with \texttt{x\_max\_ri} which is determined to be 381 cells$/mm^3 $  and 334 cells$/mm^3  $ for $ \psi_{1opt} $ and $ \psi_{2opt} $, respectively. These estimates are similar to those obtained with other methods (See Table 1). We will see in what follows how uncertainty can be quantified.
\begin{figure}[H]
	\centering
	\begin{subfigure}[b]{0.45\linewidth}
		\includegraphics[scale=0.45]{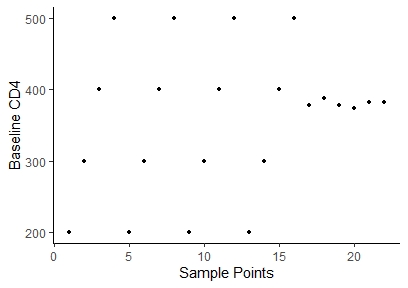}
		\caption{}
	\end{subfigure}
	\begin{subfigure}[b]{0.45\linewidth}
		\includegraphics[scale=0.45]{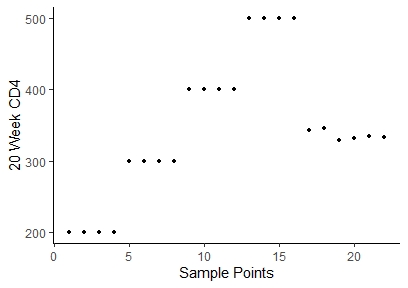}
		\caption{}
	\end{subfigure}
	\caption{Design points with additional six sequentially sampled points from estimation surface corresponding to rule "treat with zidovudine and zalcitabine when CD4 cell count is greater than $ \psi_k $ for $ k=1,2 $" (a) $ \psi_1 $ (b) $\psi_2$.}
	\label{Convergence_Assessment}
\end{figure}
 With convergence attained, the resulting posterior mean can be visualized. This can be done via the  \texttt{PostMean} function provided below.
\begin{verbatim}
#creating grid
Psi=as.matrix(expand.grid(seq(200,500,10),seq(200,500,10)))
colnames(Psi)=c("psi1","psi2")
#computing posterior mean on grid of points
estimated_y=apply(Psi,1,FUN=PostMean,GP_Object=second_fit)
\end{verbatim}
Evaluating the posterior mean on a grid of points yields Figure \ref{EmulatedSurface}. We see that it broadly resembles the other value surfaces (see Figure 2).
\begin{figure}[H]
	\centering
	\includegraphics[scale=0.23]{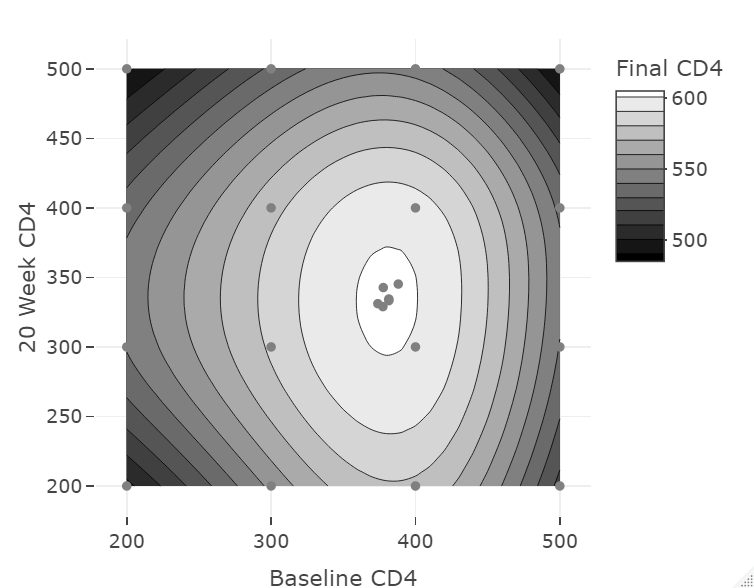}
	\caption{Emulated surface after six sequentially sampled points corresponding to the rule "treat with zidovudine and zalcitabine when CD4 cell count is greater than $ \psi_k $ for $ k=1,2 $".}
	\label{EmulatedSurface}
\end{figure}
The last step in the inferential process is to use the \texttt{FitInfer} function to quantify the uncertainty around the optimizers. Based on the convergence plots above, we chose to perform inference at 6 additionally sampled points. We perform 100 bootstraps, with 100 sampled paths in each bootstrap.
\begin{verbatim}
Location="posterior_sample.csv" #defining additional parameter for this function
#computing uncertainty around optimal DTR
Variability_Matrix=FitInfer(Design_Object=start_fit,Boot_Start=1,Boot_End=100,N=100,
				 Psi_new=Psi_new,Location=Location,Additional_Samp=6)
\end{verbatim}
Using the output of the \texttt{FitInfer} function, we compute the median and 95\% credible interval of the quantities of interest, and we obtain that $ \psi_{1opt} $ is 380 cells$/mm^3 $ (335,410), $ \psi_{2opt} $ 335 cells$/mm^3  $ (245,395), and the value at the optimum is 607 cells$/mm^3  $ (569,646). As can be seen from Table \ref{TabResults} that these estimates broadly match the results obtained with the grid search and direct modeling approaches. The credible intervals exhibit slightly more variability, but they reflect more sources of uncertainty than those that result from other methods. With these parameter settings, the \texttt{FitInfer} function takes roughly two hours to run on an \textit{Intel Core i7} processor with 16 GB of \textit{RAM}. Although priors on the covariance parameters were not used in the example analysis above,  Log-Normal priors could have been used. The code needed to specify such priors is given below. Prior parameters are set using the strategy described in section 2.4.

\begin{verbatim}
Prior_List=list(  #defining independent logged log-normal prior distributions
theta1_prior="-(log(theta1)-3.64)^2/(2*0.76^2) -log(theta1*0.76*sqrt(2*pi))",
theta2_prior="-(log(theta2)-3.64)^2/(2*0.76^2) -log(theta2*0.76*sqrt(2*pi))")

Prior_Der_List=list(  #defining derivative of logged log-normal prior distributions
theta1_der_prior="-(log(theta1)-3.64)/(0.76^2*theta1)-1/theta1",
theta2_der_prior="-(log(theta2)-3.64)/(0.76^2*theta2)-1/theta2")
\end{verbatim}

\section{Discussion}

Herein, we have examined recent Bayesian methodologies for identifying optimal dynamic treatment regimes and have used data adapted from an HIV trial to illustrate how to perform a standard DTR analysis with these methods. The \texttt{BayesDTR} package contains the \texttt{BayesMSM} function which allows users to smoothly model the value surface of regimes in a family via Bayesian dynamic MSMs with IPW estimation. These methods, through their use of a $ \mathcal{DP} $ prior with concentration parameter equal to zero, allow users to take an approach that is similar in flavour to frequentist semiparametric methods but that results in estimators that are entirely Bayesian. The function additionally allows users to perform a grid search for the optimal value, and thereby estimate the optimal treatment strategy, using a Bayesian IPW or doubly robust estimator. Given the limitations of these methods which include a potentially high computation burden (grid search methods) or vulnerability to model misspecification (parametric MSM), the package also incorporates functions that perform Gaussian process optimization to allow for the identification of optimal DTRs in conjunction with IPW and DR estimators. The \texttt{DesignFit} function in the package fits a $ \mathcal{GP} $ on an initial set of design points, and the \texttt{SequenceFit} function allows users to sequentially sample more points based on belief about where the optimum lies. Lastly, the \texttt{FitInfer} function allows users to quantify uncertainty around the optimal regime. Although this optimization takes an empirical Bayes approach, it is important that the inherent Bayesian perspective is acknowledged in this setting, as a frequentist approach, although it may work well in practice, does not acknowledge that uncertainty in the problem extends beyond aleatory uncertainty. More precisely, this means that a computer experiment as described in this paper does not have an outcome that depends on chance; the outcome will always be the same if the experiment is performed multiple times. Consequently, a frequentist framework does not accommodate the characteristics of this problem. In contrast, although the computer experiment may have a deterministic nature, there is still uncertainty about the optimal regime once the experiment is complete; it is only Bayesian methods that allow for the quantification of this uncertainty.

There are still several improvements that can be made in future versions of the package, for example introducing a function that allows the user to use an estimator for the regime value of their choosing, so as not to be limited to the ones implemented in the \texttt{BayesDTR} package. Although the illustrative analysis considered a two-stage problem, this package places no restrictions on the number of stages in the decision-problem. Additionally, as discussed, the treatment rules that can be considered with the package may involve multiple covariates per stage. Adding other methods to stabilize covariance matrices when they are near non-invertability could be beneficial, for example by adding a nugget effect. Extending the flexibility in propensity score modeling is also important, for example by allowing for regularization to select confounders as done by \citep{shortreed2017outcome} with outcome adaptive lasso.  This would require adapting these methods to allow for a Bayesian fitting procedure and to ensure correct propagation of variability to the posterior.

\section*{Acknowledgements}
DRD is supported by a doctoral fellowship from the Fonds de recherche du Qu\'ebec (FRQ), Nature et technologie. EEMM and DAS acknowledge support from Discovery Grants from the Natural Sciences and Engineering Research Council of Canada (NSERC). EEMM is a CIHR Canada Research Chair (Tier 1) in Statistical Methods for Precision Medicine and acknowledges the support of a chercheur de m\'erite career award from the FRQ, Sant\'e.

\bibliographystyle{chicago}
\bibliography{Thesis_Bibliography}

\begin{thebibliography}{}

\bibitem[\protect\citeauthoryear{Arjas and Saarela}{Arjas and
  Saarela}{2010}]{Arjas2010}
Arjas, E. and O.~Saarela (2010).
\newblock Optimal dynamic regimes: {P}resenting a case for predictive
  inference.
\newblock {\em The {I}nternational {J}ournal of {B}iostatistics\/}~{\em
  6\/}(2).

\bibitem[\protect\citeauthoryear{Artman}{Artman}{2021}]{Artman2021}
Artman, W. (2021).
\newblock {\em SMARTbayesR: Bayesian set of best dynamic treatment regimes and
  sample size in SMARTs for Binary Outcomes}.
\newblock R package version 2.0.0.

\bibitem[\protect\citeauthoryear{Austin and Stuart}{Austin and
  Stuart}{2015}]{austin2015moving}
Austin, P.~C. and E.~A. Stuart (2015).
\newblock Moving towards best practice when using inverse probability of
  treatment weighting (iptw) using the propensity score to estimate causal
  treatment effects in observational studies.
\newblock {\em Statistics in Medicine\/}~{\em 34\/}(28), 3661--3679.

\bibitem[\protect\citeauthoryear{Cain, Robins, Lanoy, Logan, Costagliola, and
  Hernán}{Cain et~al.}{2010}]{Cain2010}
Cain, L.~E., J.~M. Robins, E.~Lanoy, R.~Logan, D.~Costagliola, and M.~A.
  Hernán (2010).
\newblock When to start treatment? {A} systematic approach to the comparison of
  dynamic regimes using observational data.
\newblock {\em The {I}nternational {J}ournal of {B}iostatistics\/}~{\em
  6\/}(2).

\bibitem[\protect\citeauthoryear{Chen, Liu, Zeng, and Wang}{Chen
  et~al.}{2020}]{Chen2020}
Chen, Y., Y.~Liu, D.~Zeng, and Y.~Wang (2020).
\newblock {\em DTRlearn2: Statistical learning methods for optimizing dynamic
  treatment regimes}.
\newblock R package version 1.1.

\bibitem[\protect\citeauthoryear{Forrester, Keane, and Bressloff}{Forrester
  et~al.}{2006}]{Forrester2006}
Forrester, A.~I., A.~J. Keane, and N.~W. Bressloff (2006).
\newblock Design and analysis of "noisy" computer experiments.
\newblock {\em AIAA Journal\/}~{\em 44\/}(10), 2331--2339.

\bibitem[\protect\citeauthoryear{Frazier and Wang}{Frazier and
  Wang}{2016}]{Frazier2016}
Frazier, P.~I. and J.~Wang (2016).
\newblock Bayesian optimization for materials design.
\newblock In T.~Lookman, F.~J. Alexander, and K.~Rajan (Eds.), {\em Information
  Science for Materials Discovery and Design}, pp.\  45--75. New York:
  Springer.

\bibitem[\protect\citeauthoryear{Freeman, Browder, McGinigle, and
  Kosorok}{Freeman et~al.}{2022}]{freeman2022dynamic}
Freeman, N.~L., S.~E. Browder, K.~L. McGinigle, and M.~R. Kosorok (2022).
\newblock Dynamic treatment regime characterization via value function
  surrogate with an application to partial compliance.
\newblock {\em arXiv preprint arXiv:2212.00650\/}.

\bibitem[\protect\citeauthoryear{Ghosal and van~der Vaart}{Ghosal and van~der
  Vaart}{2017}]{ghosal2017}
Ghosal, S. and A.~van~der Vaart (2017).
\newblock {\em Fundamentals of nonparametric {B}ayesian inference}, Volume~44.
\newblock Cambridge, United Kingdom: Cambridge University Press.

\bibitem[\protect\citeauthoryear{Guan, Reich, Laber, and Bandyopadhyay}{Guan
  et~al.}{2020}]{guan2020bayesian}
Guan, Q., B.~J. Reich, E.~B. Laber, and D.~Bandyopadhyay (2020).
\newblock Bayesian nonparametric policy search with application to periodontal
  recall intervals.
\newblock {\em Journal of the American Statistical Association\/}~{\em
  115\/}(531), 1066--1078.

\bibitem[\protect\citeauthoryear{Hammer, Katzenstein, Hughes, Gundacker,
  Schooley, Haubrich, Henry, Lederman, Phair, Niu, et~al.}{Hammer
  et~al.}{1996}]{Hammer1996}
Hammer, S.~M., D.~A. Katzenstein, M.~D. Hughes, H.~Gundacker, R.~T. Schooley,
  R.~H. Haubrich, W.~K. Henry, M.~M. Lederman, J.~P. Phair, M.~Niu, et~al.
  (1996).
\newblock A trial comparing nucleoside monotherapy with combination therapy in
  {HIV}-infected adults with {CD}4 cell counts from 200 to 500 per cubic
  millimeter.
\newblock {\em New England Journal of Medicine\/}~{\em 335\/}(15), 1081--1090.

\bibitem[\protect\citeauthoryear{Henmi and Eguchi}{Henmi and
  Eguchi}{2004}]{Henmi2004}
Henmi, M. and S.~Eguchi (2004).
\newblock A paradox concerning nuisance parameters and projected estimating
  functions.
\newblock {\em Biometrika\/}~{\em 91\/}(4), 929--941.

\bibitem[\protect\citeauthoryear{Hern{\'a}n and Robins}{Hern{\'a}n and
  Robins}{2020}]{Hernan2020}
Hern{\'a}n, M.~A. and J.~M. Robins (2020).
\newblock {\em Causal inference: what if}.
\newblock Boca Raton: Chapman \& Hall/CRC.

\bibitem[\protect\citeauthoryear{Holloway, Laber, Linn, Zhang, Davidian, and
  Tsiatis}{Holloway et~al.}{2020}]{Holloway2020}
Holloway, S.~T., E.~B. Laber, K.~A. Linn, B.~Zhang, M.~Davidian, and A.~A.
  Tsiatis (2020).
\newblock {\em DynTxRegime: Methods for estimating optimal dynamic treatment
  regimes}.
\newblock R package version 4.9.

\bibitem[\protect\citeauthoryear{Hua, Mei, Zohar, Giral, and Xu}{Hua
  et~al.}{2022}]{hua2022personalized}
Hua, W., H.~Mei, S.~Zohar, M.~Giral, and Y.~Xu (2022).
\newblock Personalized dynamic treatment regimes in continuous time: a
  {B}ayesian approach for optimizing clinical decisions with timing.
\newblock {\em Bayesian Analysis\/}~{\em 17\/}(3), 849--878.

\bibitem[\protect\citeauthoryear{Huang, Allen, Notz, and Zeng}{Huang
  et~al.}{2006}]{Huang2006}
Huang, D., T.~T. Allen, W.~I. Notz, and N.~Zeng (2006).
\newblock Global optimization of stochastic black-box systems via sequential
  kriging meta-models.
\newblock {\em Journal of Global Optimization\/}~{\em 34\/}(3), 441--466.

\bibitem[\protect\citeauthoryear{Jones, Schonlau, and Welch}{Jones
  et~al.}{1998}]{Jones1998}
Jones, D.~R., M.~Schonlau, and W.~J. Welch (1998).
\newblock Efficient global optimization of expensive black-box functions.
\newblock {\em Journal of Global optimization\/}~{\em 13\/}(4), 455--492.

\bibitem[\protect\citeauthoryear{Kundu}{Kundu}{2021}]{Kundu2021}
Kundu, M.~G. (2021).
\newblock {\em LongCART: Recursive partitioning for longitudinal data and right
  censored data using baseline covariates}.
\newblock R package version 3.1.

\bibitem[\protect\citeauthoryear{Lizotte}{Lizotte}{2008}]{Lizotte2008}
Lizotte, D.~J. (2008).
\newblock {\em Practical Bayesian Optimization}.
\newblock Ph.\ D. thesis, University of Alberta, Edmonton, AB, Canada.

\bibitem[\protect\citeauthoryear{Locatelli}{Locatelli}{1997}]{Locatelli1997}
Locatelli, M. (1997).
\newblock Bayesian algorithms for one-dimensional global optimization.
\newblock {\em Journal of Global Optimization\/}~{\em 10\/}(1), 57--76.

\bibitem[\protect\citeauthoryear{{Mebane, Jr.} and Sekhon}{{Mebane, Jr.} and
  Sekhon}{2011}]{Walter2011}
{Mebane, Jr.}, W.~R. and J.~S. Sekhon (2011).
\newblock Genetic optimization using derivatives: The {rgenoud} package for
  {R}.
\newblock {\em Journal of Statistical Software\/}~{\em 42\/}(11), 1--26.

\bibitem[\protect\citeauthoryear{Murphy, van~der Laan, and Robins}{Murphy
  et~al.}{2001}]{Murphy2001}
Murphy, S.~A., M.~J. van~der Laan, and J.~M. Robins (2001).
\newblock Marginal mean models for dynamic regimes.
\newblock {\em Journal of the American Statistical Association\/}~{\em
  96\/}(456), 1410--1423.

\bibitem[\protect\citeauthoryear{Murray, Yuan, and Thall}{Murray
  et~al.}{2018}]{murray2018}
Murray, T.~A., Y.~Yuan, and P.~F. Thall (2018).
\newblock A {B}ayesian machine learning approach for optimizing dynamic
  treatment regimes.
\newblock {\em Journal of the American Statistical Association\/}~{\em
  113\/}(523), 1255--1267.

\bibitem[\protect\citeauthoryear{Myers, Rassen, Gagne, Huybrechts, Schneeweiss,
  Rothman, Joffe, and Glynn}{Myers et~al.}{2011}]{myers2011effects}
Myers, J.~A., J.~A. Rassen, J.~J. Gagne, K.~F. Huybrechts, S.~Schneeweiss,
  K.~J. Rothman, M.~M. Joffe, and R.~J. Glynn (2011).
\newblock Effects of adjusting for instrumental variables on bias and precision
  of effect estimates.
\newblock {\em American Journal of Epidemiology\/}~{\em 174\/}(11), 1213--1222.

\bibitem[\protect\citeauthoryear{Oganisian and Roy}{Oganisian and
  Roy}{2021}]{oganisian2021practical}
Oganisian, A. and J.~A. Roy (2021).
\newblock A practical introduction to {B}ayesian estimation of causal effects:
  Parametric and nonparametric approaches.
\newblock {\em Statistics in Medicine\/}~{\em 40\/}(2), 518--551.

\bibitem[\protect\citeauthoryear{O'Hagan, Kennedy, and Oakley}{O'Hagan
  et~al.}{1999}]{OHagan1999}
O'Hagan, A., M.~C. Kennedy, and J.~E. Oakley (1999).
\newblock Uncertainty analysis and other inference tools for complex computer
  codes.
\newblock In {\em Bayesian Statistics 6: Proceedings of the Sixth Valencia
  International Meeting}, pp.\  503--524. Oxford University Press.

\bibitem[\protect\citeauthoryear{Orellana, Rotnitzky, and Robins}{Orellana
  et~al.}{2010}]{Orellana2010}
Orellana, L., A.~Rotnitzky, and J.~M. Robins (2010).
\newblock Dynamic regime marginal structural mean models for estimation of
  optimal dynamic treatment regimes, part {I}: {M}ain content.
\newblock {\em The International Journal of Biostatistics\/}~{\em 6\/}(2).

\bibitem[\protect\citeauthoryear{Park and Baek}{Park and Baek}{2001}]{Park2001}
Park, J.-S. and J.~Baek (2001).
\newblock Efficient computation of maximum likelihood estimators in a spatial
  linear model with power exponential covariogram.
\newblock {\em Computers \& Geosciences\/}~{\em 27\/}(1), 1--7.

\bibitem[\protect\citeauthoryear{Picheny, Wagner, and Ginsbourger}{Picheny
  et~al.}{2013}]{Picheny2013}
Picheny, V., T.~Wagner, and D.~Ginsbourger (2013).
\newblock A benchmark of kriging-based infill criteria for noisy optimization.
\newblock {\em Structural and Multidisciplinary Optimization\/}~{\em 48\/}(3),
  607--626.

\bibitem[\protect\citeauthoryear{Robins}{Robins}{1986}]{Robins1986}
Robins, J.~M. (1986).
\newblock A new approach to causal inference in mortality studies with a
  sustained exposure period—application to control of the healthy worker
  survivor effect.
\newblock {\em Mathematical Modelling\/}~{\em 7\/}(9-12), 1393--1512.

\bibitem[\protect\citeauthoryear{Robins}{Robins}{1993}]{Robins1993}
Robins, J.~M. (1993).
\newblock Analytic methods for estimating {HIV}-treatment and cofactor effects.
\newblock In D.~G. Ostrow and R.~C. Kessler (Eds.), {\em Methodological Issues
  in AIDS Behavioral Research}, pp.\  213--287. Boston: Springer.

\bibitem[\protect\citeauthoryear{Robins, Hernan, and Brumback}{Robins
  et~al.}{2000}]{Robins2000}
Robins, J.~M., M.~A. Hernan, and B.~Brumback (2000).
\newblock Marginal structural models and causal inference in epidemiology.
\newblock {\em Epidemiology\/}~{\em 11}, 550--560.

\bibitem[\protect\citeauthoryear{Rodriguez~Duque, Stephens, and
  Moodie}{Rodriguez~Duque et~al.}{2022}]{RodriguezDuque2022b}
Rodriguez~Duque, D., D.~A. Stephens, and E.~E.~M. Moodie (2022).
\newblock Estimation of optimal dynamic treatment regimes using {G}aussian
  processes.
\newblock {\em https://doi.org/10.48550/arXiv.2105.12259\/}.

\bibitem[\protect\citeauthoryear{Rodriguez~Duque, Stephens, Moodie, and
  Klein}{Rodriguez~Duque et~al.}{2022}]{RodriguezDuque2022}
Rodriguez~Duque, D., D.~A. Stephens, E.~E.~M. Moodie, and M.~B. Klein (2022).
\newblock Semiparametric {B}ayesian inference for dynamic treatment regimes via
  dynamic regime marginal structural models.
\newblock {\em Biostatistics\/}~{\em (In Press)}.

\bibitem[\protect\citeauthoryear{Roustant, Ginsbourger, and Deville}{Roustant
  et~al.}{2012}]{Roustant2012}
Roustant, O., D.~Ginsbourger, and Y.~Deville (2012).
\newblock Dicekriging, diceoptim: Two {R} packages for the analysis of computer
  experiments by kriging-based metamodelling and optimization.
\newblock {\em Journal of Statistical Software\/}~{\em 51\/}(1), 1--55.

\bibitem[\protect\citeauthoryear{Rubin}{Rubin}{1981}]{Rubin1981}
Rubin, D.~B. (1981).
\newblock The {B}ayesian bootstrap.
\newblock {\em The Annals of Statistics\/}~{\em 9\/}(1), 130--134.

\bibitem[\protect\citeauthoryear{Saarela, Arjas, Stephens, and Moodie}{Saarela
  et~al.}{2015}]{Saarela2015a}
Saarela, O., E.~Arjas, D.~A. Stephens, and E.~E.~M. Moodie (2015).
\newblock Predictive {B}ayesian inference and dynamic treatment regimes.
\newblock {\em Biometrical Journal\/}~{\em 57\/}(6), 941--958.

\bibitem[\protect\citeauthoryear{Saarela, Belzile, and Stephens}{Saarela
  et~al.}{2016}]{Saarela2016}
Saarela, O., L.~R. Belzile, and D.~A. Stephens (2016).
\newblock A {B}ayesian view of doubly robust causal inference.
\newblock {\em Biometrika\/}~{\em 103\/}(3), 667--681.

\bibitem[\protect\citeauthoryear{Saarela, Stephens, Moodie, and Klein}{Saarela
  et~al.}{2015}]{Saarela2015}
Saarela, O., D.~A. Stephens, E.~E.~M. Moodie, and M.~B. Klein (2015).
\newblock On {B}ayesian estimation of marginal structural models.
\newblock {\em Biometrics\/}~{\em 71\/}(2), 279--288.

\bibitem[\protect\citeauthoryear{Santner, Williams, Notz, and Williams}{Santner
  et~al.}{2018}]{Santner2018}
Santner, T.~J., B.~J. Williams, W.~Notz, and B.~J. Williams (2018).
\newblock {\em The design and analysis of computer experiments\/} (Second ed.).
\newblock New York: Springer.

\bibitem[\protect\citeauthoryear{Shortreed and Ertefaie}{Shortreed and
  Ertefaie}{2017}]{shortreed2017outcome}
Shortreed, S.~M. and A.~Ertefaie (2017).
\newblock Outcome-adaptive lasso: variable selection for causal inference.
\newblock {\em Biometrics\/}~{\em 73\/}(4), 1111--1122.

\bibitem[\protect\citeauthoryear{Stephens, Nobre, Moodie, and Schmidt}{Stephens
  et~al.}{2022}]{Stephens2021}
Stephens, D.~A., W.~S. Nobre, E.~E.~M. Moodie, and A.~M. Schmidt (2022).
\newblock Causal inference under mis-specification: adjustment based on the
  propensity score.
\newblock {\em https://doi.org/10.48550/arXiv.2201.12831\/}.

\bibitem[\protect\citeauthoryear{Tsiatis, Davidian, Holloway, and
  Laber}{Tsiatis et~al.}{2019}]{tsiatis2019dynamic}
Tsiatis, A.~A., M.~Davidian, S.~T. Holloway, and E.~B. Laber (2019).
\newblock {\em Dynamic treatment regimes: Statistical methods for precision
  medicine}.
\newblock New York: Chapman and Hall/CRC.

\bibitem[\protect\citeauthoryear{Walker}{Walker}{2010}]{Walker2010a}
Walker, S.~G. (2010).
\newblock Bayesian nonparametric methods: motivation and ideas.
\newblock In N.~L. Hjort, C.~Holmes, P.~M{\"u}ller, and S.~G. Walker (Eds.),
  {\em Bayesian {N}onparametrics}, Chapter~1. New York: Cambridge University
  Press.

\bibitem[\protect\citeauthoryear{Wallace, Moodie, Stephens, Simoneau, and
  Schulz}{Wallace et~al.}{2020}]{Wallace2020}
Wallace, M.~P., E.~E.~M. Moodie, D.~A. Stephens, G.~Simoneau, and J.~Schulz
  (2020).
\newblock {\em DTRreg: DTR estimation and inference via g-estimation, dynamic
  WOLS, Q-learning, and dynamic weighted survival modeling (DWSurv)}.
\newblock R package version 1.7.

\bibitem[\protect\citeauthoryear{Williams and Rasmussen}{Williams and
  Rasmussen}{2006}]{williams2006gaussian}
Williams, C.~K. and C.~E. Rasmussen (2006).
\newblock {\em Gaussian processes for machine learning}, Volume~2.
\newblock MIT press Cambridge, MA.

\bibitem[\protect\citeauthoryear{Xiao, Abrahamowicz, and Moodie}{Xiao
  et~al.}{2010}]{Xiao2010}
Xiao, Y., M.~Abrahamowicz, and E.~E.~M. Moodie (2010).
\newblock Accuracy of conventional and marginal structural {C}ox model
  estimators: a simulation study.
\newblock {\em The International Journal of Biostatistics\/}~{\em 6\/}(2).

\bibitem[\protect\citeauthoryear{Xu, M{\"u}ller, Wahed, and Thall}{Xu
  et~al.}{2016}]{Xu2016}
Xu, Y., P.~M{\"u}ller, A.~S. Wahed, and P.~F. Thall (2016).
\newblock Bayesian nonparametric estimation for dynamic treatment regimes with
  sequential transition times.
\newblock {\em Journal of the American Statistical Association\/}~{\em
  111\/}(515), 921--950.

\bibitem[\protect\citeauthoryear{Zhao, Kosorok, and Zeng}{Zhao
  et~al.}{2009}]{zhao2009}
Zhao, Y., M.~R. Kosorok, and D.~Zeng (2009).
\newblock Reinforcement learning design for cancer clinical trials.
\newblock {\em Statistics in Medicine\/}~{\em 28\/}(26), 3294--3315.

\bibitem[\protect\citeauthoryear{Zhao, Zeng, Rush, and Kosorok}{Zhao
  et~al.}{2012}]{zhao2012}
Zhao, Y., D.~Zeng, J.~Rush, and M.~R. Kosorok (2012).
\newblock Estimating individualized treatment rules using outcome weighted
  learning.
\newblock {\em Journal of the American Statistical Association\/}~{\em
  107\/}(499), 1106--1118.

\end{thebibliography}

\newpage
\appendix
{\Large \textbf{Appendix}}
\section{Covariance Details}
For a covariance matrix $ \mathbbm{K} $ computed via a covariance function $ \mathbbm{k}(\psi_i,\psi_j) $  and parameterized by $ \eta_f=(\theta_f, \sigma^2_f) $, the $Mat\acute{e}rn_{5/2}$ covariance function between two regime indices $ \psi_i, \psi_j $ is given by:
\begin{equation*}
	\mathbbm{k}(\psi_i,\psi_j)=\sigma^2_{f}\prod_{d=1}^{D}\left(1+\frac{\sqrt{5}|\psi_{id}-\psi_{jd}|}{\theta_{fd}}+
	\frac{5(\psi_{id}-\psi_{jd})^2}{3\theta_{fd}^2} \right)\exp\left(\frac{-\sqrt{5}|\psi_{id}-\psi_{jd}|}{\theta_{fd}}\right),
\end{equation*}
where $ D $ is the dimension of $ \psi $ and $ \psi_{id} $ and $ \theta_{fd} $ are the $ d $th entries in the $ \psi_i $  and $ \theta_f $ vectors, respectively. $ \sigma^2_f $ scales the correlation function to yield the covariance.

\section{Parameters for BayesMSM function}
We begin by examining the parameters in the \texttt{BayesMSM} function.
\begin{verbatim}
BayesMSM(PatID,Data,Outcome_Var,Treat_Vars,Treat_M_List,Outcome_M_List,
        MSM_Model,G_List,Psi,Bayes=TRUE,DR=FALSE,Normalized=FALSE,B=100,Bayes_Seed=1)
\end{verbatim}

\noindent\textbf{Aim 1: Marginal Structural Model}
\begin{itemize}
\item \textbf{Required}: \texttt{PatID}, \texttt{Data}, \texttt{Outcome\_Var}, \texttt{Treat\_Vars}, \texttt{Treat\_M\_List}, \texttt{MSM\_Model}, \texttt{G\_List}, \texttt{Psi}
\item \textbf{Optional}: \texttt{Bayes}, \texttt{B}, \texttt{Bayes\_Seed}
\item \textbf{Unavailable}: \texttt{Normalized}
\end{itemize}

\textbf{Aim 2: Grid-Search IPW Estimator}
\begin{itemize}
	\item \textbf{Required}: \texttt{PatID}, \texttt{Data}, \texttt{Outcome\_Var}, \texttt{Treat\_Vars}, \texttt{Treat\_M\_List},  \texttt{G\_List}, \texttt{Psi}
	\item \textbf{Optional}: \texttt{Bayes}, \texttt{Normalized}, \texttt{B}, \texttt{Bayes\_Seed}
\end{itemize}

\textbf{Aim 3: Grid-Search Double Robust Estimator}
\begin{itemize}
	\item \textbf{Required}: \texttt{PatID}, \texttt{Data}, \texttt{Outcome\_Var}, \texttt{Treat\_Vars}, \texttt{Treat\_M\_List}, \texttt{Outcome\_M\_List},  \texttt{G\_List}, \texttt{Psi}, \texttt{DR=TRUE}
	\item \textbf{Optional}: \texttt{Bayes}, \texttt{Normalized}, \texttt{B}, \texttt{Bayes\_Seed}
\end{itemize}

\newpage
\section{Parameters for Gaussian Process Functions}
We now examine the parameters required to perform the Gaussian Process optimization.
\begin{verbatim}
DesignFit(PatID,Data,Outcome_Var,Treat_Vars,Treat_M_List,Outcome_M_List,
	  Normalized=TRUE,DR=FALSE,G_List,Psi,
	  Covtype,Numbr_Samp,IthetasU,IthetasL,Likelihood_Limits=NA,
	  Prior_List=NULL,Prior_Der_List=NULL)
\end{verbatim}

\begin{itemize}
	\item \textbf{Required}:  \texttt{PatID}, \texttt{Data}, \texttt{Outcome\_Var}, \texttt{Treat\_Vars}, \texttt{Treat\_M\_List}, \texttt{G\_List}
	 \texttt{Numbr\_Samp}, \texttt{IthetasU}, \texttt{IthetasL}, \texttt{Covtype}
	 \begin{itemize}
	 	\item Note: The default is to use the normalized IPW estimator.
	 \end{itemize}
	
	\item \textbf{Optional}: \texttt{Outcome\_M\_List}, \texttt{Normalize}, \texttt{DR}, \texttt{Likelihood\_Limits}, \texttt{Prior\_List}, \texttt{Prior\_Der\_List}
	\begin{itemize}
		\item Note: The double robust estimator can be used by specifying the \texttt{Outcome\_M\_List} and \texttt{DR} parameters.
	\end{itemize}
\end{itemize}

\begin{verbatim}
SequenceFit(Previous_Fit,Additional_Samp,Control_Genoud=list())
\end{verbatim}
\begin{itemize}
	\item \textbf{Required}: \texttt{Previous\_Fit}, \texttt{Additional\_Samp}, \texttt{Control\_Genoud}
	\begin{itemize}
	\item  Note that in particular the \texttt{Control\_Genoud} function requires the \texttt{Domain} element. We have demonstrated the use of this parameter in the main paper (section 3.2).
	\end{itemize}
\end{itemize}

\begin{verbatim}
FitInfer(Design_Object,Boot_Start,Boot_End,Psi_new,N,Location,
		 Additional_Samp,Control_Genoud=list())
\end{verbatim}
\begin{itemize}
	\item \textbf{Required}: \texttt{Design\_Object}, \texttt{Boot\_Start}, \texttt{Boot\_End}, \texttt{Psi\_new}, \texttt{N}, \texttt{Additional\_Samp}, \texttt{Control\_Genoud}
	\item \textbf{Optional}: \texttt{Location}
\end{itemize}

\end{document}